\documentclass[11pt, epsfig]{article}
\usepackage{epsfig, amsmath, amssymb, amsthm, times}
\def\theequation{\arabic{section}.\arabic{equation}}
\renewcommand{\theequation}{\thesection.\arabic{equation}}
\newtheorem {thm}{Theorem}[section]
\newtheorem {lem}[thm]{Lemma}

\theoremstyle{defintion}
\newtheorem {df}[thm]{Definition}

\theoremstyle{remark}
\newtheorem{rem}[thm]{Remark}

\def\qed{\hfill $\Box$ \hfill \\}

\def\p{\partial}
\def\lbl{\label}
\def\be{\begin{equation}}
\def\ee{\end{equation}}
\def\lbl{\label}
\def\Tb{\mathcal{T}}
\def\la{\lambda}
\def\pf{\noindent{\em Proof }}
\def\var{{\rm var}}
\def\cov{{\rm cov}}
\def\E{{\mathbb E}}
\def\P{{\mathbb P}}
\def\Re{{\mathbb R}}

\def\eps{\epsilon}

\def\xt{\xi_t}
\def\het{\theta_t}
\def\x1{\xi^{(1)}_t}
\def\t1{\theta_t^{(0)}}

\title{Bursting oscillations induced by small noise}
\author{
Pawel Hitczenko and
Georgi S. Medvedev
\thanks{
Department of Mathematics, Drexel University, 3141 Chestnut Street,
Philadelphia, PA 19104, {\tt phitczen@math.drexel.edu, medvedev@drexel.edu}
}
\thanks{to appear in SIAM J. Appl. Math.; submitted December 25, 2007; accepted  November 7, 2008}
}
\begin{document}
\maketitle
\begin{abstract}
We consider a model of a square-wave bursting neuron residing in the regime of tonic spiking. Upon
introduction of small stochastic forcing, the model generates irregular bursting. The statistical properties
of the emergent bursting patterns are studied in the present work. In particular, we identify two principal
statistical regimes associated with  the noise-induced bursting. In the first case, (type I) bursting
oscillations are created mainly due to the fluctuations in the fast subsystem. In the alternative scenario,
type II bursting, the random perturbations in the slow dynamics play a dominant role. We propose two classes
of randomly perturbed slow-fast systems that realize type I and type II scenarios. For these models, we
derive the Poincare maps. The analysis of the linearized Poincare maps of the randomly perturbed systems
explains the distributions of the number of spikes within
one burst and reveals their dependence on the small and control parameters present in the models. The
mathematical analysis of the model problems is complemented by the numerical experiments with a generic
Hodgkin-Huxley type model of a bursting neuron. 
\end{abstract}

\section{Introduction}
\begin{figure}
\begin{center}
{\bf a}\epsfig{figure=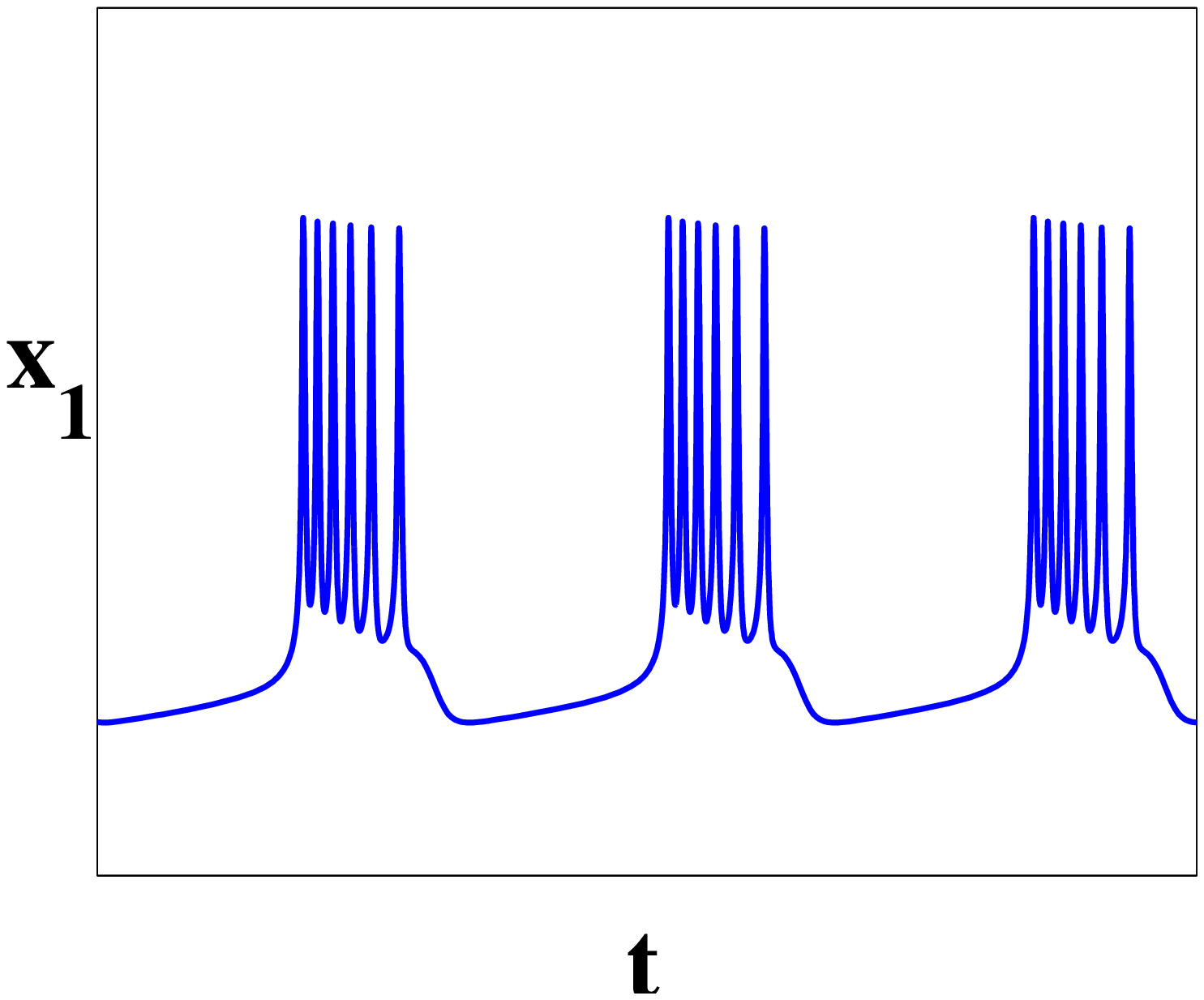, height=2.2in, width=3.0in, angle=0}\quad
{\bf b}\epsfig{figure=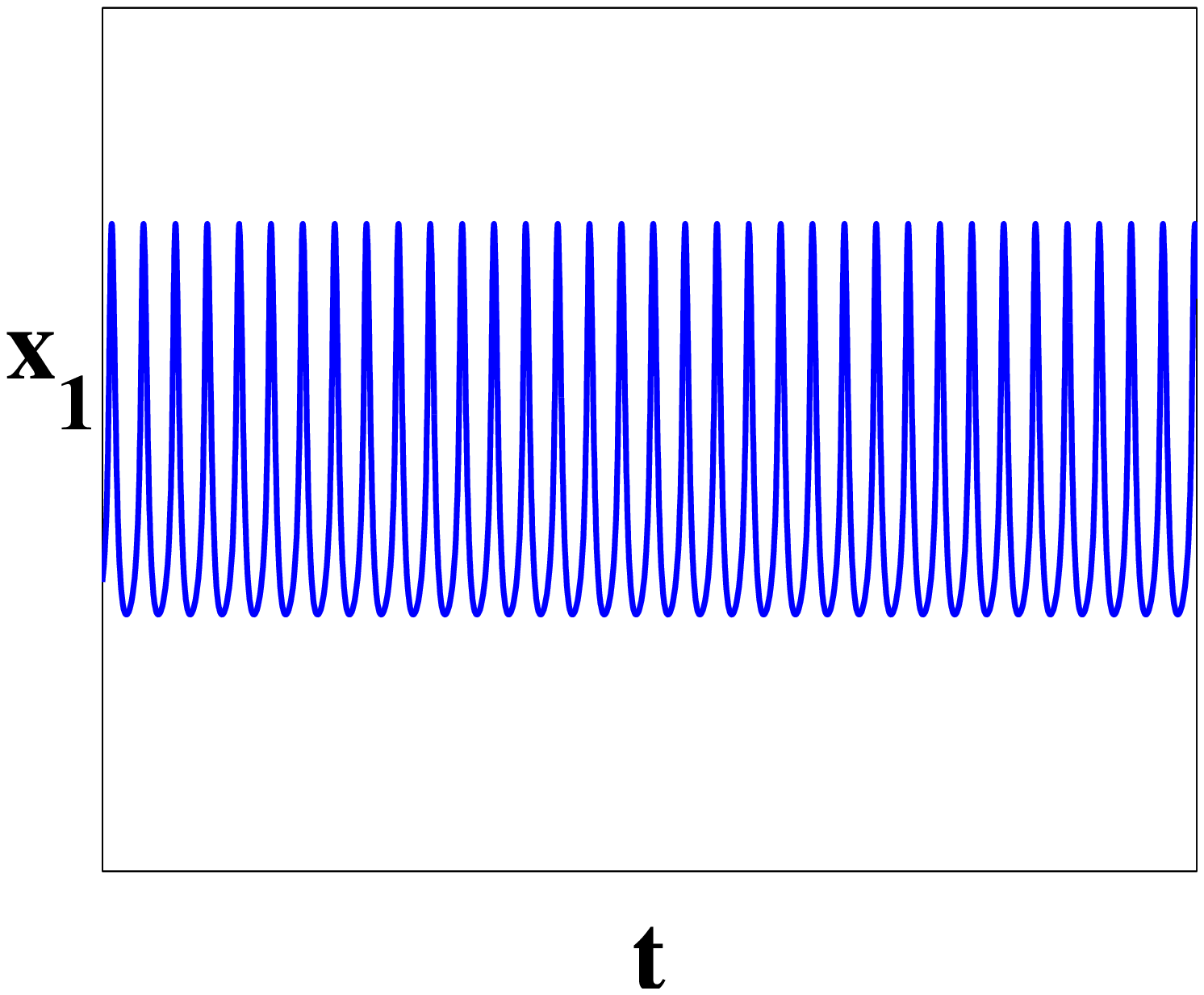, height=2.2in, width=3.0in, angle=0}
\end{center}
\caption{ The dynamical patterns generated by a model of of a square-wave
bursting neuron (\ref{1.1}) and (\ref{1.2}): (a) periodic bursting and 
(b) tonic spiking.  
}
\label{f.0}
\end{figure}

Differential equation models of excitable cells often include small random terms 
to reflect the unresolved or
poorly understood aspects of the problem or to account for intrinsically 
stochastic factors \cite{BC, chow_white, CCI, DVM, fox, fox_lu, LH, smith, rowat_elson, SRT, wrk}. 
In addition,
many neuronal models also exhibit multistability \cite{RE89, IZH07}. 
In systems with multiple stable states, noise may induce
transitions between different attractors in the system dynamics, thus, creating qualitatively new dynamical
regimes, that are not present in the deterministic system. In the present paper, we study this situation for
a class of square-wave bursting models of excitable cell membranes. 
This class includes many conductance-based models of 
excitable cell membranes. Here we just mention the model of a pancreatic
$\beta-$cell \cite{chay85, CR}, models of neurons in various central pattern
generators such as those involved in insect locomotion \cite{GH04a}, control of the
heartbeat in a leech \cite{hill}, and respiration in mammals \cite{BBRTW, BRS}, to name
a few. 
These models, as well as the underlying biological systems, exhibit characteristic bursting 
patterns of the voltage time series: clusters of fast spikes 
alternating with pronounced periods of quiescence (Fig. \ref{f.0}a).
For introduction to bursting, examples and bibliography, we refer the reader to 
\cite{IZH07, lee_terman, rinzel87, RE89}. 
The dynamical patterns generated by the conductance-based models typically depend sensitively 
on parameters.
For example, models of square-wave bursting neurons often exhibit both bursting and spiking behaviors for 
different values of parameters (see Fig. \ref{f.0}a,b). In many relevant experiments, the transition from 
spiking to bursting is achieved by changing the injected current. 
In the present paper, we consider a model of 
a square-wave bursting neuron in the regime of tonic spiking (Fig. \ref{f.0}b). We show that a small noise can transform 
spiking patterns into irregular (noise-induced) bursting patterns and
describe two distinct mechanisms for generating
noise-induced bursting. In the first scenario, bursting oscillations are triggered by the fluctuations in the
fast subsystem. We refer to this mechanism as type I bursting. In contrast, 
the bursting dynamics in type II
scenario are driven by the random motion along the slow manifold. For each of these cases, we describe the
statistical properties of the emergent bursting patterns and characterize them in terms of the small and
control parameters present in the model.

Noise-induced phenomena have received considerable attention in 
the context of neuronal modeling (see, e.g.,
\cite{BC, chow_white, CCI, LH, rowat_elson, smith, SRT, wrk}). 
A representative example is given by a $2D$ excitable system
perturbed by the white noise of small intensity \cite{BC}. In the presence of noise and under
certain general conditions, a typical
trajectory occasionally leaves the basin of attraction (BA) of the stable equilibrium and makes a large
excursion in the phase plane of the deterministic system before returning to a small neighborhood of the
stable fixed point (Fig. \ref{f.-1}a). This gives rise to irregular spiking (Fig. \ref{f.-1}b).  
The properties of the noise-induced spiking and stochastic resonance type effects arising in the context of 
the perturbed FitzHugh-Nagumo model have been
considered in \cite{BC, chow_white, CCI, DVM} (see also \cite{BG, FR01, FR01a, FW} for the mathematical analysis of more general classes of related phenomena in randomly perturbed slow-fast systems). 
In the present paper, we study a related mechanism for irregular bursting.
Specifically, we consider a class of models of square-wave bursting neurons:
\begin{eqnarray}\lbl{1.1}
\dot x &=& f(x,y),\\
\lbl{1.2}
\dot y &=& \epsilon g(x,y),\quad x=(x^1, x^2)^T\in\Re^2,\; y\in\Re^1,
\end{eqnarray}
where $f$ and $g$ are smooth functions and $0<\epsilon\ll 1$ is a small parameter. 
We refer to (\ref{1.1}), where $y$ is treated as a parameter, as a fast subsystem.
It is formally obtained from (\ref{1.1}) and (\ref{1.2}) 
by setting $\epsilon=0$. We assume that the fast subsystem  has a family of
stable limit cycles and that of stable equilibria for $y$ in a certain interval $y\in (y_{sn}, y_{bp})$ (see
Fig. \ref{f.1}a). The additional assumptions on (\ref{1.1}) and (\ref{1.2}), which are explained in Section
2, imply that for small $\epsilon>0$, (\ref{1.1}) and (\ref{1.2}) has a stable limit cycle as shown in Fig.
\ref{f.1}c. In the presence of noise, a typical trajectory of the randomly perturbed system will occasionally
leave the BA of the limit cycle of the deterministic system to make an excursion along the curve of
equilibria of the degenerate system, E  (see Fig. \ref{f.2}a). Thus, in analogy to the $2D$ FitzHugh-Nagumo 
model (Fig. \ref{f.-1}a), noise
transforms spiking dynamics into irregular bursting. We refer to the latter as noise-induced bursting.
 In both examples above, irregular spiking (Fig. \ref{f.-1}a) or bursting patterns 
(Fig. \ref{f.2}a,b) are created due to the escape of a trajectory
 of the randomly perturbed system from the BA of a stable fixed point in the case
 of spiking or of that of the stable limit cycle in the case of bursting. The statistics of the first exit
 times can then be related to the properties of the emergent firing patterns such as the frequency of spiking
 or the distribution of the number of spikes within one burst. Compared to the analysis
of the irregular spiking in the randomly perturbed
 FitzHugh-Nagumo model (Fig.~\ref{f.-1}) , the analysis of the noise-induced bursting faces several additional challenges due to
 the fact that in
 the latter case one has to consider the exit problem for the trajectories near a stable limit cycle as opposed
 to those near a stable equilibrium in the former case. The structure of the BA of the limit cycle combined
 with the slow-fast character of the vector field determines the main features of the resultant bursting
 patterns.  The description of the principal statistical regimes associated with the noise-induced bursting
 is the focus of the present paper.
\begin{figure}
\begin{center}
{\bf a}\epsfig{figure=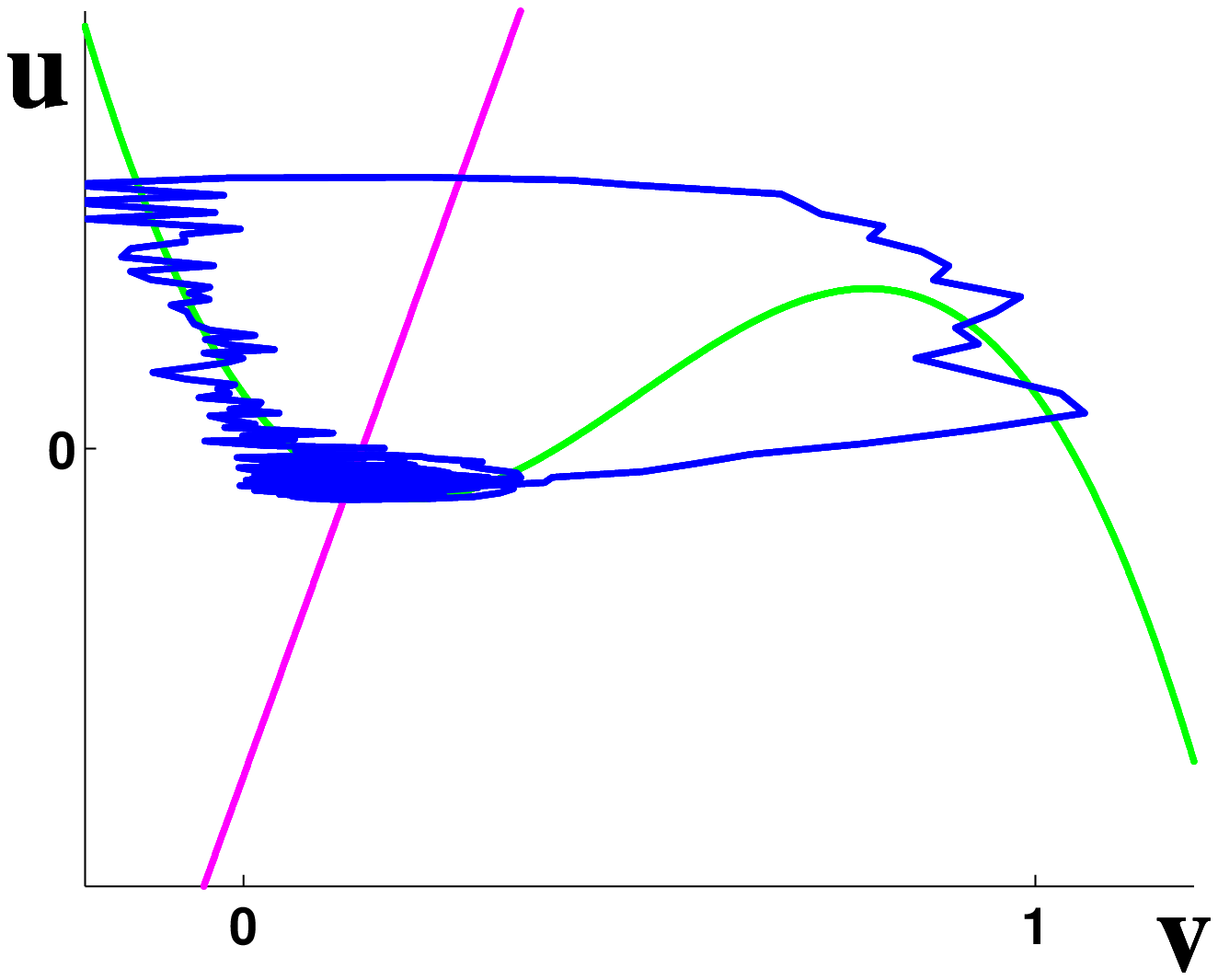, height=2.2in, width=3.0in, angle=0} \quad
{\bf b}\epsfig{figure=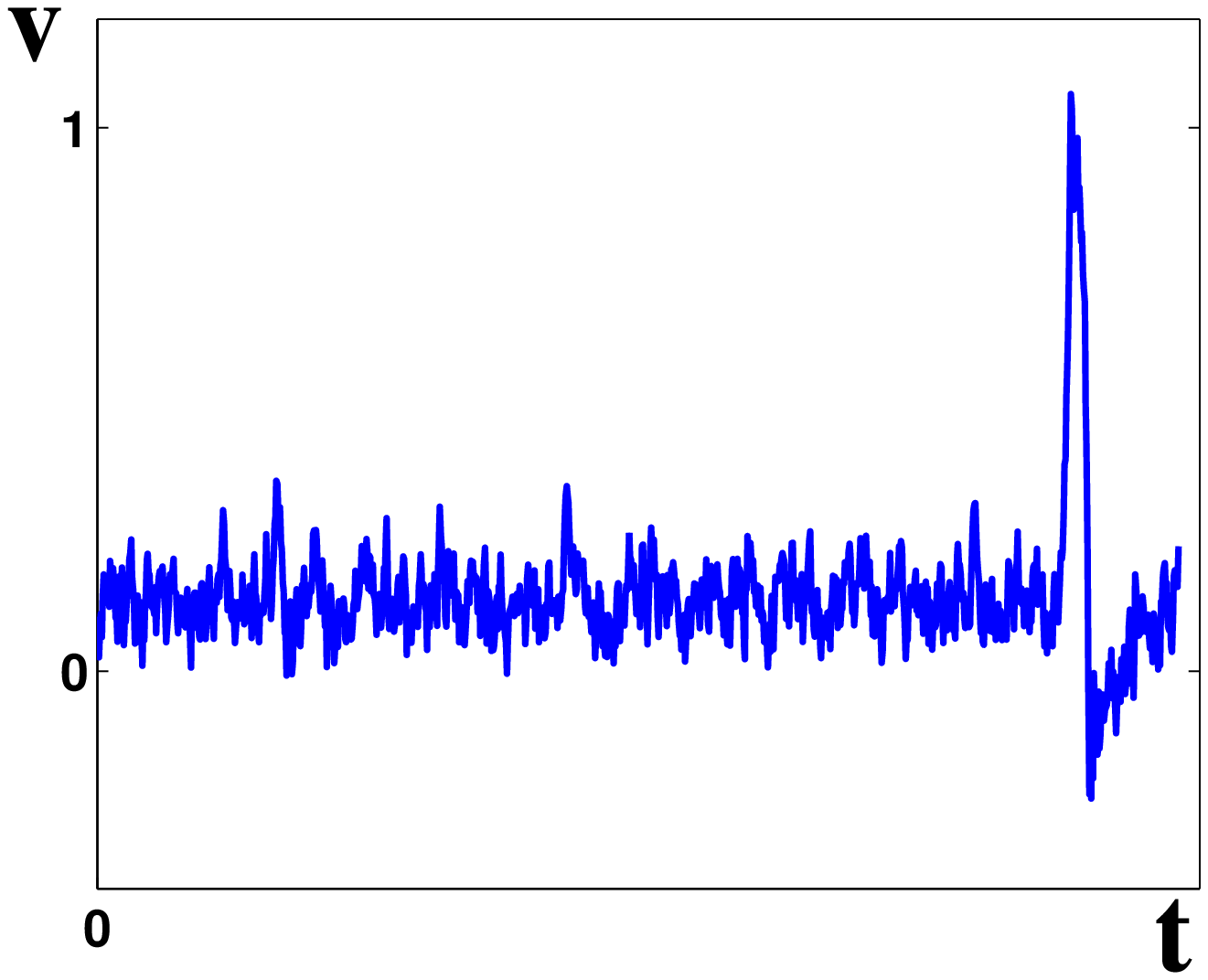, height=2.2in, width=3.0in, angle=0}
\end{center}
\caption{(a) A phase-plane trajectory of the randomly perturbed FitzHugh-Nagumo model
in excitable regime (see \cite{BC} for the model description and the parameter values). 
(b) The time series corresponding to the phase plot in (a).
}\lbl{f.-1}
\end{figure}

There are general mathematical approaches for analyzing exit problems
for stochastic processes generated by randomly perturbed differential equations 
such as (\ref{1.1}) and (\ref{1.2}): the Wentzell-Freidlin theory of large deviations \cite{FW} and 
the geometric theory for randomly perturbed slow-fast systems due to Berglund and Gentz \cite{BG}. 
In this paper we study the vector fields arising in the context of bursting. 
The specialized structure of this class of problems allows us to keep the analysis of the present paper self-contained 
and avoid using more technical methods, which are necessary for analyzing more general situations.
Our analytical approach is based on the reduction of a randomly perturbed differential equation
model to the Poincare map and studying the exit problems for the trajectories of the discrete system.
Using maps is quite natural in the context of bursting due to the intrinsic discreteness of bursting patterns 
imposed by the presence of spikes. Reductions to maps have been very useful for analyzing bursting
dynamics in a variety of deterministic models \cite{chay85, medvedev06, medvedev05, MC, rinzel_troy82a}. 
As follows from the results of the present paper, the first return maps also provide a very convenient 
and visual representation for the 
mechanism underlying noise-induced bursting. In particular, we show that the distributions of spikes in one
burst in many cases are effectively determined by $1D$ linear randomly perturbed maps. We develop a set of
probabilistic techniques for analyzing the dynamics of randomly perturbed $1D$ and $2D$ linear maps such as
those arising in the analysis of bursting. 
The special structure of this class of problems, which is motivated by the applications to bursting
affords a more direct and simpler analysis than the treatment of more general classes of random linear
maps found in the literature \cite{kesten, goldie,johnkotz, vervaat}. 

The outline of the paper is as follows. In section 2, we formulate
our assumptions on the deterministic system. We then present the
preliminary numerical results, motivating our formulation of the
randomly perturbed models at the end of this section.
Specifically, we distinguish two types of the noise-induced
bursting. {\it Type I}  bursting is generated due to the
fluctuations predominantly in the fast subsystem, while
{\it type II} bursting is induced by variability mainly in the slow
variable. Accordingly, we introduce two types of models that
generate type I and type II bursting patterns. Section 3 develops
a set of probabilistic techniques, which will be needed for the
analysis of the first return maps for the randomly perturbed
differential equation models. We first analyze a simple linear
map with an attracting slope and small additive Gaussian
perturbations in Section 3.2. Due to the simple structure of the map,
we obtain very explicit characterization of the
first exit times for this problem. The analysis of this first
relatively simple example provides the guidelines for the more
complex cases dealt in Sections 3.3-3.5.
Section 4 contains the definition and the construction
of the Poincare map for the type I randomly perturbed model
introduced in Section 2.
The $2D$ Poincare map is decomposed into two $1D$ maps for
the fast and slow subsystems, which are constructed in Sections 4.2 and 4.3
respectively. In Section 4.4, we apply the results of Section 3 to the
linearization of the Poincare map to derive the distributions of
the first exit times. The latter are interpreted as the distributions of
the number of spikes in one burst. 
In Sections 4.5, we outline the modifications necessary to
cover type II models. Since the analysis for type II
models closely follows the lines of that for type I models, we omit most
of the details. Finally, the numerical experiments in Section 5 are
designed to illustrate our theory. 

\section{The model}
\setcounter{equation}{0}
In the present section, we introduce the model to be studied in the remainder of this
paper. We start by formulating our assumptions on the deterministic model and then
describe the random perturbation.

\subsection{The deterministic model}

We consider slow-fast system  (\ref{1.1}) and (\ref{1.2}) in $\Re^3$ with one {\it slow}
variable. The {\it fast} subsystem associated with (\ref{1.1}) and (\ref{1.2}) is
obtained by sending $\epsilon\to 0$ in (\ref{1.2}) and treating $y$ as a parameter:
\be\lbl{1.3}
\dot x = f(x,y).
\ee

\begin{figure}
\begin{center}
{\bf a}\epsfig{figure=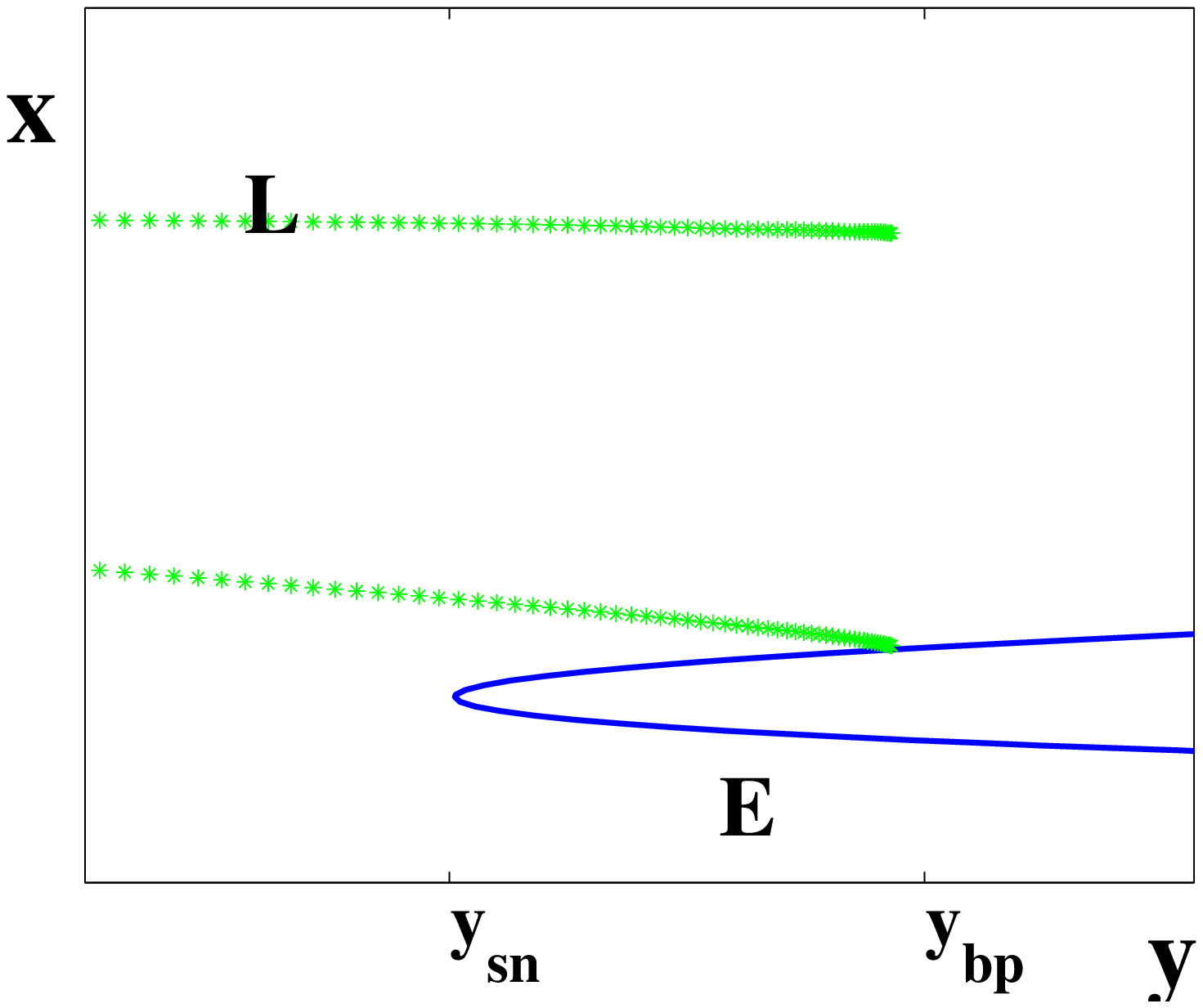, height=2.0in, width=2.0in, angle=0}
{\bf b}\epsfig{figure=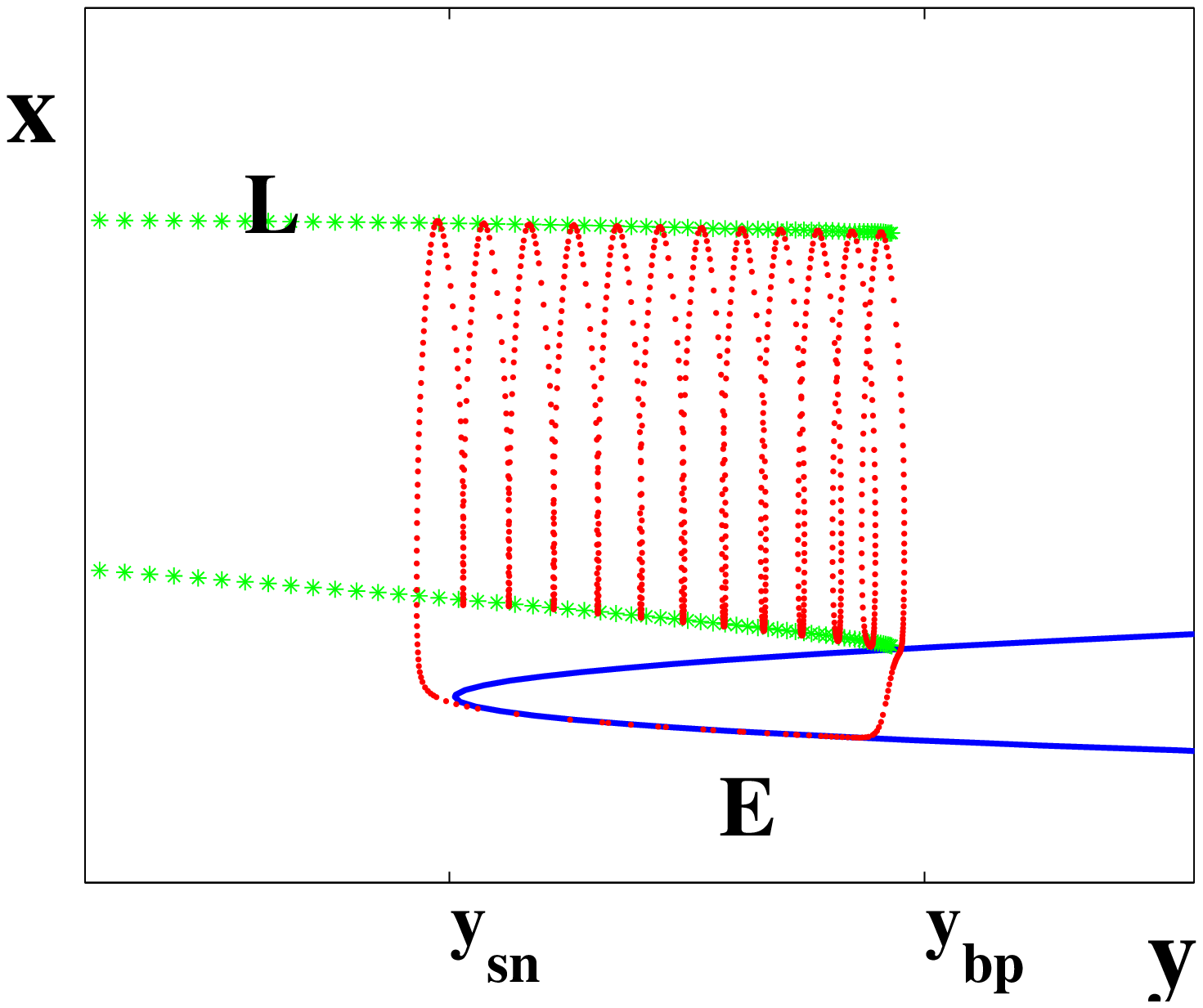, height=2.0in, width=2.0in, angle=0}
{\bf c}\epsfig{figure=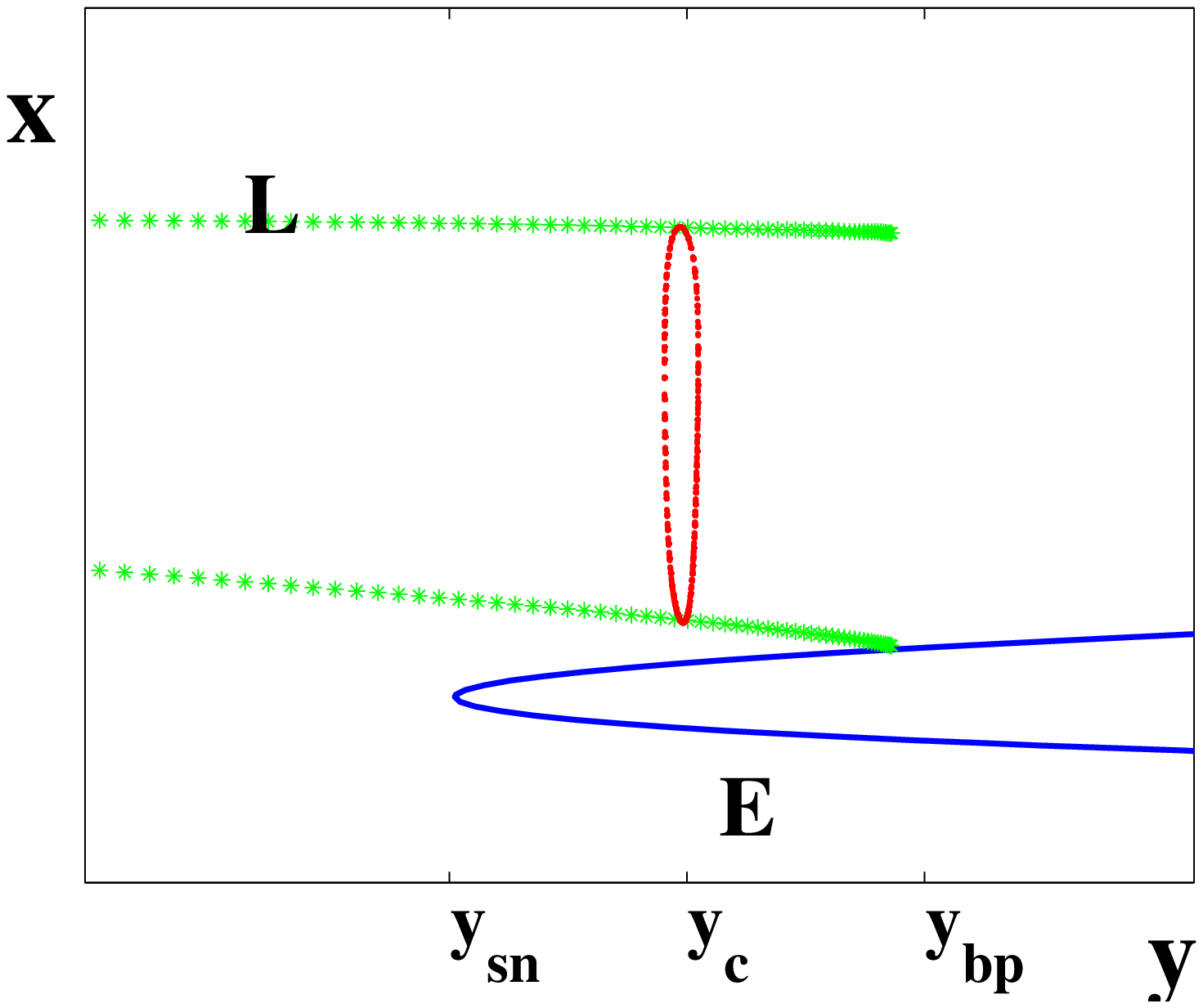, height=2.0in, width=2.0in, angle=0}
\end{center}
\caption{
(a) The bifurcation diagram of the fast subsystem (\ref{1.3}).
$L$ denotes a cylinder foliated by the stable periodic orbits.
The lower branch of the parabolic curve $E$
is composed of stable equilibria of the fast subsystem
(see Fig. \ref{f.4}b for the  plot of a representative phase 
plane of the fast subsystem for $y\in (y_{sn},y_{bp})$).
(b,c) Periodic trajectories of the full system (\ref{1.1}) and
(\ref{1.2}) are superimposed on the bifurcation diagram of the fast subsystem.
Assumptions (SE) and (SB) (see the text) result in a bursting limit cycle
plotted in red in (b), while (SS) yields spiking (c).
}
\label{f.1}
\end{figure}

Under the variation of $y$, the fast subsystem has the bifurcation structure
as shown schematically in Fig. \ref{f.1}a. Specifically,
we rely on the following assumptions:
\begin{description}
\item[(PO)]
There exists $y_{bp}\in\Re$ such that for each $y<y_{bp}$, Equation (\ref{1.3}) has an exponentially
stable limit cycle of period $\mathcal{T}(y)$:
\be\lbl{1.4}
L(y)=\{x=\phi(s,y): \;0\le s<\Tb(y)\}.
\ee
The family of the limit cycles, $L=\bigcup_{y<y_{bp}}L(y)$, forms a cylinder in $\Re^3$
(Fig. \ref{f.1}a).

\item[(EQ)]
There is a branch of asymptotically stable equilibria of (\ref{1.3}),
$E=\left\{x=\psi(y):\; y>y_{sn}\right\}$, which terminates at
a saddle-node bifurcation at $y=y_{sn}< y_{bp}$ (Figure \ref{f.1}a).

\item[(LS)] For each $y\in\Re$, the $\omega-$limit set of almost all trajectories of (\ref{1.3})
belongs to $L(y)\bigcup \{\psi(y)\}$.
\end{description}

\begin{rem}\lbl{F2}
At $y=y_{bp}$, $L$, either terminates or $L(y_{bp}+0)$ looses stability.
We do not specify the
type of the bifurcation at $y=y_{bp}$. It may be, for instance, a homoclinic bifurcation
as shown in Fig. \ref{f.1}a, or a saddle-node bifurcation of limit cycles \cite{GH}.
\end{rem}
Having specified the assumptions on the bifurcation structure of the fast
subsystem, we turn to the slow dynamics. The geometric theory for
singularly perturbed systems implies the existence of the
exponentially stable locally invariant manifolds $E_\epsilon$ and
$L_\epsilon$, which are $O(\epsilon)$ close to
$E\bigcap\{(x,y): y>y_{sn}+\delta\}$ and $L\bigcap\{(x,y): y<y_{bp}-\delta\}$,
respectively, for arbitrary fixed $\delta>0$ and sufficiently small $\epsilon>0$
\cite{fenichel, jones}.
Manifolds $E_\epsilon$ and $L_\epsilon$ are called {\it slow manifolds}.
For small $\eps>0$, the dynamics of (\ref{1.1}) and (\ref{1.2})
on the slow manifolds is approximated by
\begin{eqnarray}\lbl{1.6}
L_\epsilon:&\qquad\qquad\qquad\dot y =\epsilon G(y),& y<y_{bp}-\delta,\\
\lbl{1.7}
E_\epsilon:&\qquad\qquad\qquad\dot y =\epsilon g(\psi(y),y),& y>y_{sn}+\delta,
\end{eqnarray}
where
\be\lbl{1.8}
G(y)={1\over\Tb(y)} \int_0^{\Tb(y)} g\left(\phi(s),y\right) ds.
\ee

We distinguish two types of the asymptotic behavior of solutions of (\ref{1.1}) and (\ref{1.2}):
{\em bursting} and {\em spiking} (see Fig. \ref{f.0}). 
The following conditions on the slow subsystem yield bursting.

\noindent
For some $c>0$ independent of $\eps$,
\begin{description}
\item[(SE)]
\be\lbl{1.9}
  g(\psi(y),y)<-c \quad \mbox{for}\quad y>y_{sn},
\ee
\item[(SB)]
\be\lbl{1.9a}
G(y)>c \quad \mbox{for}\quad y<y_{bp}.
\ee
\end{description}
Under these assumptions, for sufficiently small $\eps>0$ a typical trajectory of (\ref{1.1}) and (\ref{1.2})
consists of the alternating segments
closely following $L_\eps$ and $E_\eps$ and fast transitions between them
(see Fig. \ref{f.1}b).
For detailed discussions of the geometric construction of 'bursting` periodic
orbits, we refer the reader
to \cite{lee_terman, rinzel87}.
To obtain spiking,  we substitute (SB) with
\begin{description}
\item[(SS)] $G(y)$ has a unique simple zero at $y=y_c\in (y_{sn},y_{bp})$:
\be\lbl{1.10}
G(y_c)=0 \quad\mbox{and}\quad G^\prime(y_c)<0.
\ee
\end{description}
In this case, the asymptotic behavior of solutions follows from the following theorem due to
Pontryagin and Rodygin:
\begin{thm}\cite{PR}
If $\epsilon>0$ is sufficiently small, (\ref{1.1}) and (\ref{1.2}) has a unique exponentially stable
limit cycle $L_\eps(y_c)$ of period $\Tb(y_c)+O(\epsilon)$ lying in an $O(\epsilon)$ neighborhood of
$L(y_c)$, provided (SS) holds.
\end{thm}
Almost all trajectories of (\ref{1.1}) and (\ref{1.2}) are attracted by the limit cycle
lying in an $O(\epsilon)$ neighborhood of $L(y_c)$. This mode of behavior is called spiking
(see Fig. \ref{f.1}c and Fig. \ref{f.1}b). In the remainder of this paper we assume (SS), in addition, to
(PO), (EQ), (LS), and (SE).
\begin{figure}
\begin{center}
{\bf a}\epsfig{figure=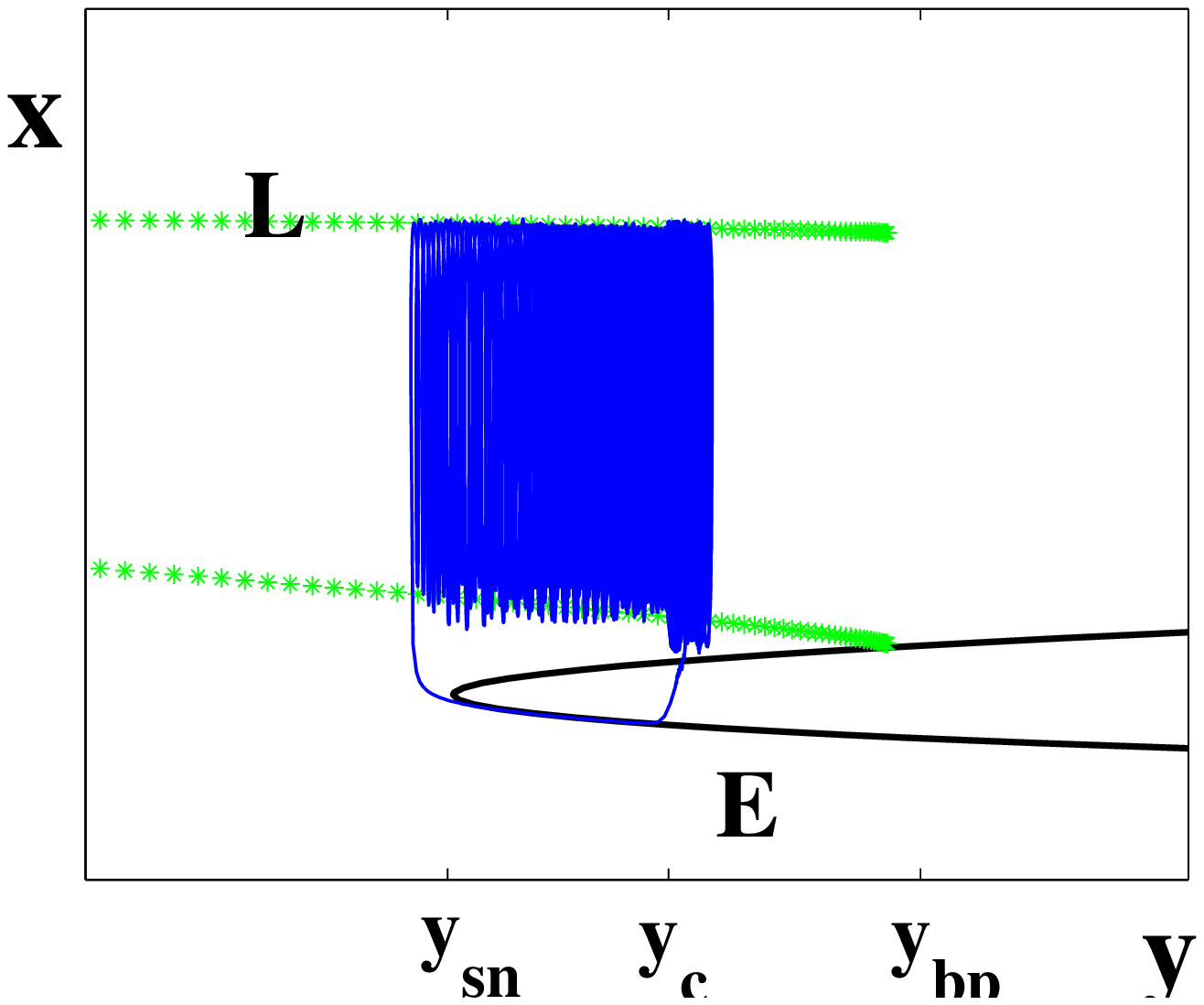, height=2.0in, width=2.5in, angle=0}
{\bf b}\epsfig{figure=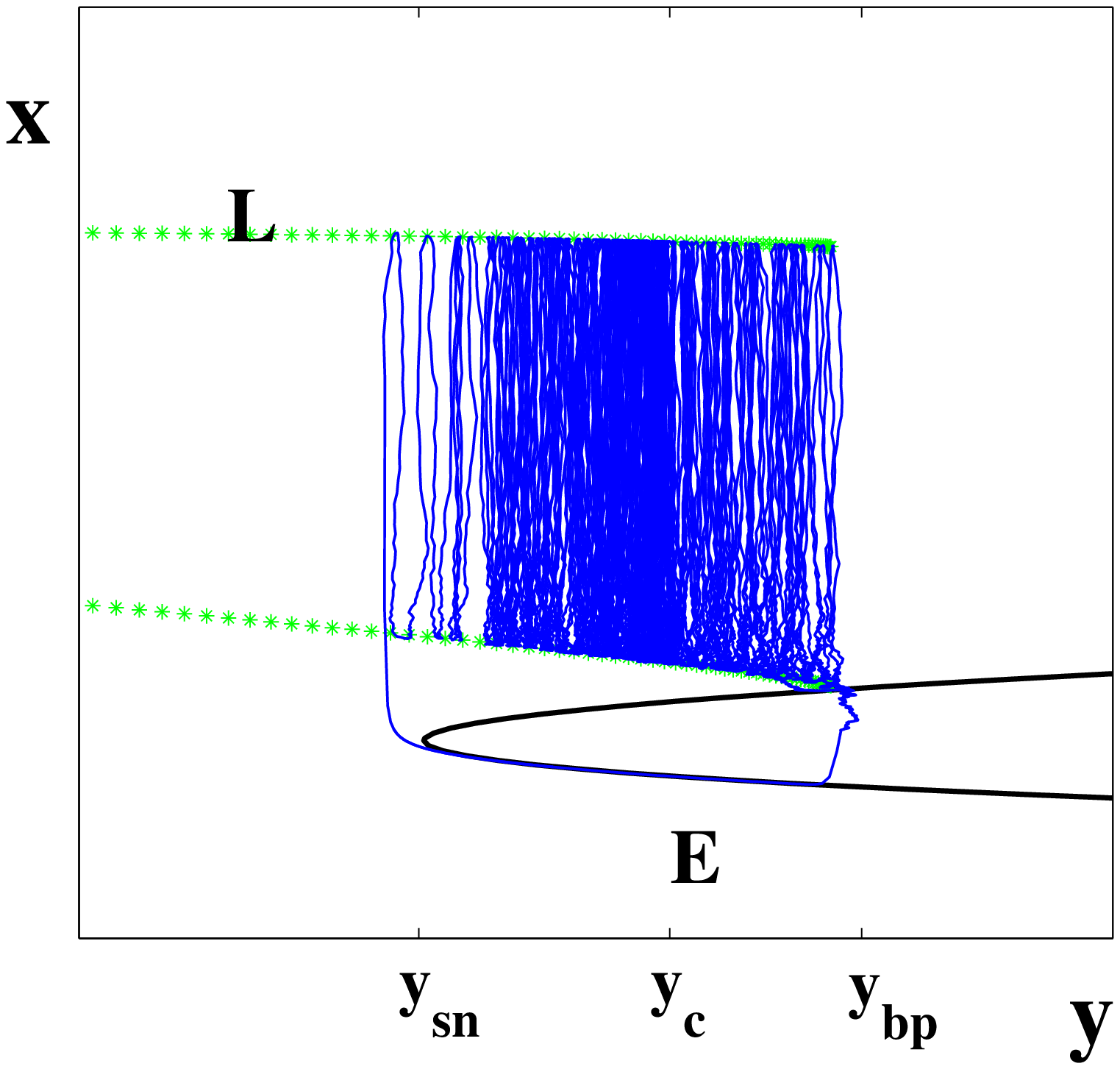, height=2.0in, width=2.5in, angle=0} \\
{\bf c}\epsfig{figure=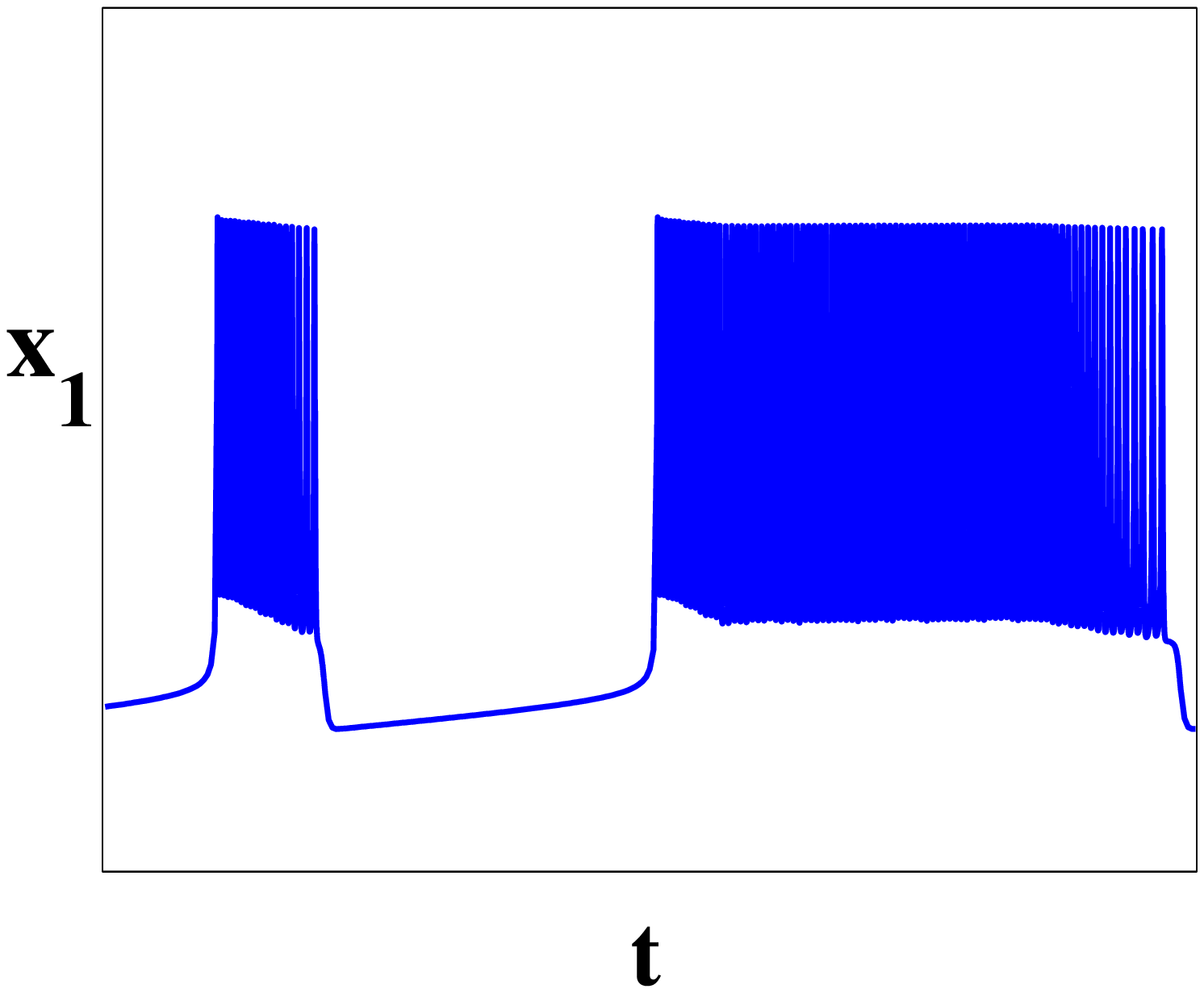, height=2.0in, width=2.5in, angle=0}
{\bf d}\epsfig{figure=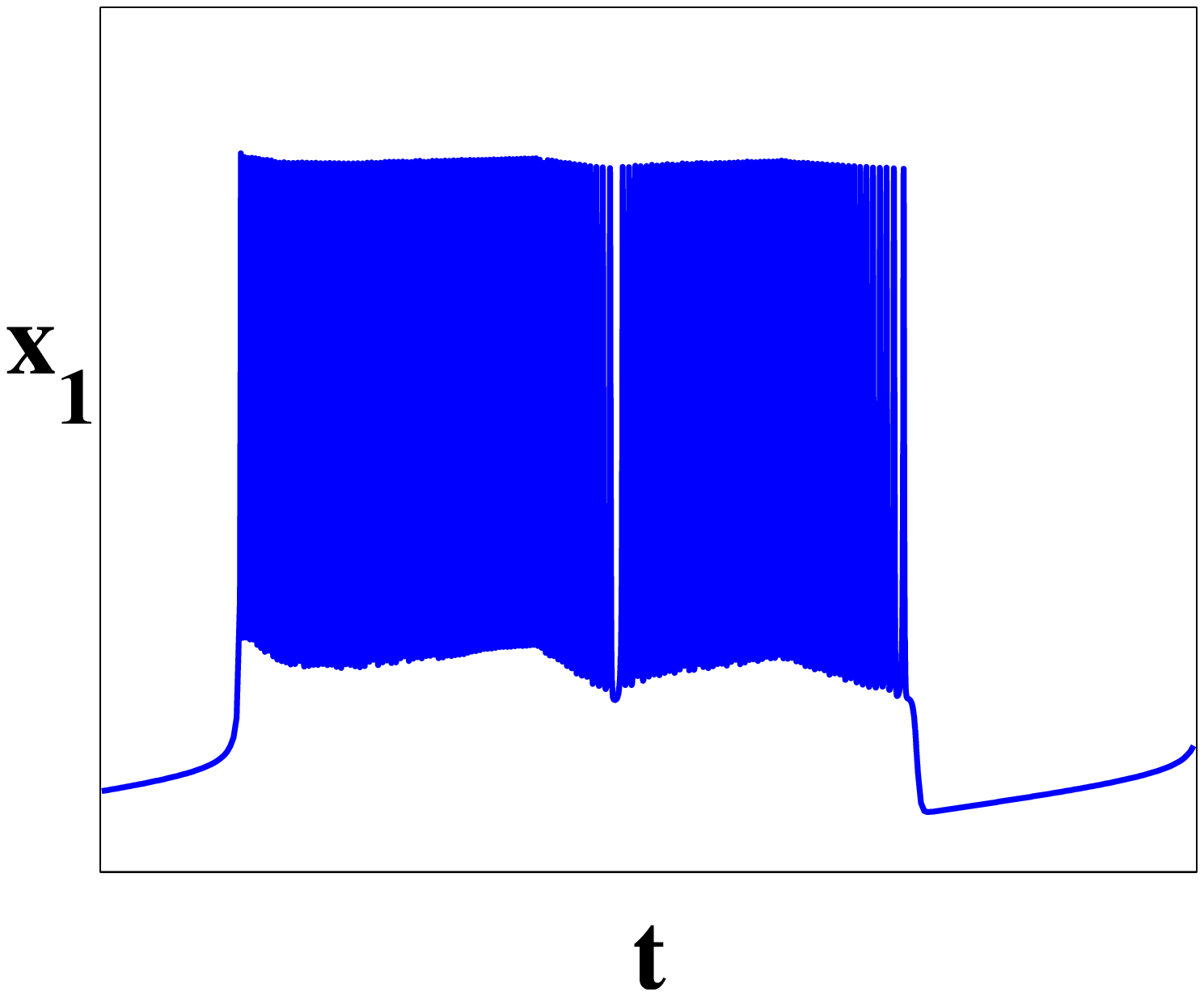, height=2.0in, width=2.5in, angle=0}
\end{center}
\caption{Noise-induced bursting. (a) A trajectory of the randomly perturbed system
is shown in the phase space of the frozen system (\ref{1.1}), (\ref{1.2}) with
$\epsilon=0$. The trajectory leaves the basin of $L(y_c)$ mainly due to the fluctuations
in the fast plane. This is characteristic to {\it type I} bursting. An alternative {\it type II}
scenario is shown in plot (b), where the fluctuations in the slow direction dominate in
the mechanism of escape from the basin of the stable limit cycle. The trajectory in (b) samples
a wide region of $L$ and leaves a neighborhood of $L$ near the right boundary, $y\approx y_{bp}$;
while that in (a) remains near $L(y_c)$ most of the time and jumps down near $y\approx y_c$.
The differences translate into the distinctive features of the generic time series of the
bursting patterns generated via type I or type II mechanisms shown in plots (c) and (d) respectively.
Note that the longer burst in (c) has a typical square-wave form (roughly, determined by $L(y_c)$),
while the burst shown in (d) exhibits more variability due to the drifting of the trajectory along $L$.
}
\label{f.2}
\end{figure}

\subsection{The randomly perturbed models}
In this subsection, we provide a heuristic description of the effects
of the random perturbations on the dynamics of (\ref{1.1}) and (\ref{1.2}). To study
these effects quantitatively, at the end of this section, we propose two randomly
perturbed models.

Suppose the trajectories of (\ref{1.1}) and (\ref{1.2}) experience weak stochastic
forcing, such that the perturbed trajectories represent well-defined stochastic processes
and are close to the trajectories of (\ref{1.1}) and (\ref{1.2}) on finite intervals
of time. Since the trajectories of the unperturbed system remain in a small neighborhood
of $L(y_c)$ (possibly after short transients), we expect that in the presence of noise
the trajectories will occasionally leave the BA of $L(y_c)$
and after making a brief excursion along $E$ will return back to the vicinity
of $L(y_c)$. Therefore, under random perturbation the system can
exhibit bursting dynamics, while the underlying deterministic system is in the spiking
regime. We refer to this mode of behavior as {\em noise-induced} bursting.
Our goal is to describe typical statistical regimes associated with the noise-induced
bursting and to relate them to the structure of (\ref{1.1}) and (\ref{1.2}) and to
the properties of the stochastic forcing. 
To illustrate the implications of the structure of the  deterministic vector field
for the bursting patterns that it produces under random perturbations, we refer to the
following numerical examples. Note that the BA of $L(y_c)$ naturally extends along the
cylinder of periodic orbits $L$ (Fig. \ref{f.1}c). The escape from the BA
of $L(y_c)$ can be dominated
by the fluctuations along $L$ or by those in the transverse plane. These two possibilities
are shown in Fig. \ref{f.2}. The trajectory shown in Fig. \ref{f.2}a spends most
of the time near $L(y_c)$  and leaves its BA due to the fluctuations in the fast
subsystem. We refer to this scenario as {\em type I} escape. Alternatively, the
trajectory shown in Fig. \ref{f.2}b travels a good deal along $L$ before the escape
and exits from the BA near $y=y_{bp}$. This mechanism is dominated by the slow
dynamics. We refer to this scenario as {\em type II} escape. These mechanisms of escape
translate into distinct features of the resultant bursting patterns.
First, note that since in type I and type II scenarios, the transition
from spiking to quiescence typically takes place at $y\approx y_c$ and $y\approx y_{bp}$
respectively, by (\ref{1.2}) and (EQ), the corresponding interburst intervals
are approximately equal to
$$
IBI\approx \eps^{-1}\int^{y_{sn}}_{\hat{y}} {dy\over g\left(\psi(y),y\right)},
\quad\mbox{where}\quad
\left\{\begin{array}{cc}
\hat{y}=y_{c}, &\;\mbox{type I},\\
\hat{y}=y_{bp}, &\;\mbox{type II}.
\end{array}
\right.
$$
In addition, we expect that the interspike intervals (ISIs) within one
burst in type I scenario are localized about $\Tb(y_c)$,
since the trajectory of the randomly perturbed system in the active phase of bursting
spends most of the time near $L(y_c)$.
In type II bursting patterns, ISIs are expected to have more variability,
since the trajectories sample a wider range of ISIs during their
excursions along $L$. Perhaps, a more pronounced distinction between these two types
of bursting patterns lies in the degree of the variability of the spikes in one burst.
Most of the spikes forming a burst
in type I pattern are generated by (\ref{1.3}) with $y\approx y_c$ and, therefore,
are similar in shape (Fig. \ref{f.2}c). In contrast, spikes in type II scenario
are subject to more variability and the bursting patterns typically have
ragged shape (Fig. \ref{f.2}d).

To study type I and type II noise-induced bursting patterns it is convenient to
consider two types of models. {\em Type I model} incorporates random forcing in
the fast subsystem:
\begin{eqnarray}\lbl{0.1}
\dot x_t &=& f\left(x_t, y_t\right)+\sigma p\dot w_t, \\
\lbl{0.2}
\dot y_t &=& \epsilon g(x_t, y_t),
\end{eqnarray}
while, in {\em type II model} the slow subsystem is forced
\begin{eqnarray}
\lbl{0.1a}
\dot x_t &= &f\left(x_t, y_t\right), \\
\lbl{0.2a}
\dot y_t& =& \eps \left(g(x_t, y_t)+\sigma q \dot w_t\right).
\end{eqnarray}
Here, $0<\sigma\ll 1$, $p(x,y)=\left(p^1(x,y),p^2(x,y)\right)^T$ and $q(x,y)$  are differentiable
functions; $\dot w_t$ stands for the white noise, i.e. a generalized derivative
of the Wiener process.

\section{The randomly perturbed maps}
\setcounter{equation}{0}

In this section, we develop probabilistic tools needed for the analysis of
randomly perturbed systems (\ref{0.1})-(\ref{0.2a}).
The number of spikes in one burst is a natural random variable  associated with
the noise-induced bursting. It is commonly used in the experimental studies of
bursting and we shall adopt it for characterizing irregular bursting patterns in
this work.
In Section 4, we will show that the number of spikes in one burst is represented by a stopping time
(more precisely, the level exceedance time) of a discrete random process, the Poincare map
of the randomly perturbed system (\ref{0.1})-(\ref{0.2a}). In preparation for
the analysis of the linearized Poincare map in Section 4, in the present section we 
study certain stochastic linear difference equations. 
The equations of this form equations have been considered in the literature before.
The study was initiated by Kesten~\cite{kesten} who considered multidimensional case
(in which the coefficients of the stochastic equations are random matrices).
Subsequent work focused mostly on the $1D$ case. We refer the reader
to the papers \cite{goldie,vervaat}, which contain representative results,
examples of applications, and further references. There is also a review paper 
\cite{eg}, unfortunately not easily accessible.
The convergence properties  of the solutions that we will need could be deduced from a general
theory of stochastic difference equations. However,   the results in the literature are
often stated in the most general form and some of the proofs are rather involved.
We will be dealing with special cases that are much easier to justify. For this reason,
and also to keep the paper self-contained we will include the proofs of the needed results.

\subsection{Geometric random variables}
We begin by recalling the necessary properties of geometric random variables (RVs).
Recall that $Y$ is a geometric RV with parameter $p$, $0<p<1$ if
\be\lbl{4.2}
\P\left(Y=k\right)=p(1-p)^{k-1}, \quad k\ge 1.
\ee
We refer the reader to \cite[Chapter~5]{johnkotz} for the review of the properties of geometric
distributions and their applications.
In particular, the following characterization of geometric RVs is classical.
% We include a short proof.
 \noindent
 \begin{lem} \lbl{lem:geom} Let $Y$ be a RV with values in the set of positive integers.
  $Y$ is a geometric  with parameter $p$,  $0<p<1$, iff
 \begin{equation}\label{geom}
 \P(Y=n)=p\P(Y\ge n),\quad n\ge1.\end{equation}
 \end{lem}
%  \pf:\;
%  The necessity of (\ref{geom}) follows directly from (\ref{4.2}).
%  We prove that (\ref{geom}) gives also a sufficient condition by induction.
%  For $k=1$, (\ref{4.2}) follows from (\ref{geom}), since $\P(Y\ge1)=1$.
% Assume that (\ref{4.2}) holds for $k=n-1$, $n>1$.
%  By (\ref{geom}), we have
%  $$
%  \P(Y=n)=p\left(1-\P(Y< n)\right)=p-p\sum_{k=1}^{n-1}\P(Y=k).$$
%  Writing a similar expression for $\P(Y=n-1)$ and by subtracting the latter from the
% former, we have
%  $$\P(Y=n)-\P(Y=n-1)=-p\sum_{k=1}^{n-1}\P(Y=k)-(-p\sum_{k=1}^{n-2}\P(Y=k))
%  =-p\P(Y=n-1).$$
%  Hence, $\P(Y=n)=(1-p)\P(Y=n-1)$ and $\P(Y=n)=p(1-p)^{n-1}$, as required.
%  \hfill\qed

Lemma~\ref{lem:geom} motivates the following definition:

\begin{df}\lbl{def:asgeom}
Let $Y$ be a random variable with values in the set of positive integers and let $0<p<1$. 
We say that $Y$ is asymptotically geometric with parameter $p$ if
\begin{equation}\lbl{asgeom}
\lim_{n\to\infty} \frac{\P\left(Y=n\right)}{\P(Y\ge n)}=p.
\end{equation}
\end{df}

\subsection{The randomly perturbed map: additive perturbation}

Consider
\be\lbl{5.1}
Y_{n}=\lambda Y_{n-1} +\varsigma r_n,\quad n\ge1,
\ee
where
$r_1,r_2,\dots$ are independent identically distributed (IID) copies of the standard normal RV, and $Y_0$ is a real number. 
We will use $N(\mu,\eta^2)$ notation for a normal RV with mean $\mu$, variance $\eta^2$, and probability density function 
given by
$$\frac1{\sqrt{2\pi}\eta}\exp\left\{-\frac{(x-\mu)^2}{2\eta^2}\right\},\quad -\infty<x<\infty.$$ 
We will also let $Z$  denote a generic  $N(0,1)$ RV and we will write
$$\Phi(x):=\frac1{\sqrt{2\pi}}\int\limits_{-\infty}^xe^{-t^2/2}dt,$$
for its   distribution function. 
For a given $h>0$, let
$$
\tau=\inf\{ k\ge 1:\ Y_k>h \}.
$$
\begin{thm}\lbl{thm:addpert}
Let
\be\lbl{5.3}
\varepsilon\in (0,1), \quad \lambda=1-\varepsilon,\quad
\beta^2={\varsigma^2\over\varepsilon(2-\varepsilon)}, \quad\mbox{and}\quad h-Y_0>0.
\ee
Then for sufficiently small $\varsigma>0$, $\tau$ is
asymptotically geometric RV with parameter
\be\lbl{5.3a}
p=\frac1
{\sqrt{2\pi}}
{\beta\over
h\Phi(h/\beta)} \exp\left\{-{h^2\over 2\beta^2}\right\}\left(1+O\left(\frac\varsigma\varepsilon\right)^2\right).
\ee
% where
% $$\Phi(x)=\frac1{\sqrt{2\pi}}\int\limits_{-\infty}^xe^{-t^2/2}dt,$$
% is the  distribution function of an  $N(0,1)$ r.v. .
\end{thm}
% \begin{rem}\lbl{rm:sigma-epsilon}
% Note that $\varsigma=o(\epsilon)$ as follows from the proof.
% \end{rem}

\noindent
We precede the proof of the theorem with the auxiliary
\begin{lem}\lbl{lem:eta}
For $n\ge1,$ $Y_n$ is a normal RV with
\begin{equation}\lbl{5.2}
\E~Y_n=\lambda^nY_0 \quad\mbox{and}\quad
\var~Y_n={\varsigma^2\left(1-\lambda^{2n}\right)\over 1-\lambda^2}=:\beta_n^2.
\end{equation}
In particular, 
$$Y_n\stackrel{d}{\longrightarrow} Y\stackrel d=N(0, \beta^2),
$$
where $\stackrel{d}{\longrightarrow}$ (and $\stackrel{d}=$) denote the convergence (equality) in distribution.
\end{lem}

\pf\; (Lemma~\ref{lem:eta}): 
The statements in (\ref{5.2}) are verified by a straightforward calculation. 
The rest follows, because $\E~Y_n\to 0$ and $\beta_n\to\beta$.

\pf\; (Theorem~\ref{thm:addpert}): Let $Y_k^*=\max\{Y_j:\ 1\le j\le k\}$, $k\ge1$. Then
\begin{eqnarray} \nonumber
\P(\tau=n+1)&=&\P(Y_{n+1}>h,Y_n^*\le h)=
\P(Y_{n+1}>h|Y_n^*\le h)\P(Y_n^*\le h)\\
\nonumber
&=&\P(Y_{n+1}>h|Y_n\le h, Y_{n-1}\le h,\dots, Y_0\le h)\P(\tau\ge n+1)\\
\lbl{markov}
&=&\P(Y_{n+1}>h|Y_n\le h)\P(\tau\ge n+1).
\end{eqnarray}
In the last equality, we used the fact that $\{Y_n\}$ is a Markov process which is clear from (\ref{5.1}).
By (\ref{markov}),
\be\lbl{5.4}
p_n:={\P(\tau=n+1)\over\P(\tau\ge n+1)}=
\P\left(Y_{n+1}>h\left| Y_n \le h\right.\right)=\frac{\P\left(Y_{n+1}>h, Y_n\le h\right)}{\P\left(Y_n\le h\right)}.
\ee
In accordance with Definition~\ref{def:asgeom}, we need to show that $\{p_n\}$ converges and
to estimate the limit.
By Lemma~\ref{lem:eta}, %$(Y_n-\lambda^nY_0)/\beta_n$ is $N(0,1)$ and, thus,
$$
\P\left( Y_n\le h\right)%=\P\left(Z\le\frac{h-\lambda^nY_0}{\beta_n}\right)
\longrightarrow \Phi(h/\beta),\;\;
\mbox{as}\;\; n\to\infty.
$$
Next, we turn to estimating the numerator
in (\ref{5.4}). We have
\begin{eqnarray*}
Q_n &:=& \P\left(Y_{n+1}>h, Y_n\le h\right)=\P\left(\lambda Y_n+\varsigma r_{n+1}>h, Y_n\le h\right)\\
    & \to& \P\left(\lambda Y+\varsigma Z>h,Y\le h\right)=:Q,
\end{eqnarray*}
where $Z$ is standard normal, $Y$ is $N(0,\beta^2)$ and they are independent. This follows from  Lemma~\ref{lem:eta} and the fact that $r_{n+1}$ is $N(0,1)$ and is independent of $Y_n$. 
$Q$ is the probability that a 2D Gaussian vector is in the region 
$[h,\infty)\times(-\infty,h]$. There are several ways of estimating this probability. 
We take the following, elementary approach. Let $X=h-Y$ so that $X$ is 
$N\left(h, \beta^2\right)$ and is independent of $Z$.
Then
$$%\be\lbl{5.8}
Q=\P\left( Z>{\varepsilon\over\varsigma}h +{1-\varepsilon\over\varsigma}X, X\ge0\right)
={1\over\sqrt{2\pi} \beta} \int_0^\infty \P \left(Z> {\varepsilon h+(1-\varepsilon)s \over \varsigma}\right) e^{-(s-h)^2\over 2\beta^2} ds.
$$
By the well--known asymptotics (see \cite[Ch.~VII, Lemma~2 and Sec.~7, Problem~1]{fel})
\be\lbl{tail}
\P(Z>u)=1-\Phi(u)=\frac1{\sqrt{2\pi}}\frac{e^{-\frac{u^2}{2}}}u\left(1+O\left(\frac1{u^2}\right)\right),\quad u>0.
\ee
%%{\tt Q: Is (\ref{tail}) valid for large $u$?} \\
Hence, for sufficiently small $\varsigma>0\; (\varsigma\ll \varepsilon)$, we have
\be\lbl{5.8aa}
Q\approx  {1\over2\pi} {\varsigma\over\beta}
\int^{\infty}_0
\frac{\exp\left\{-\frac12\left(\frac{(\varepsilon h+(1-\varepsilon )s)^2}{\varsigma^2}+\frac{(s-h)^2}{\beta^2}\right)\right\}}
{\varepsilon h+(1-\varepsilon )s}ds.
\ee
Since
$$\frac{(\varepsilon h+(1-\varepsilon)s)^2}{\varsigma^2}
+ \frac{(s-h)^2}{\beta^2}
=
\frac{(s-\varepsilon h)^2}{\varsigma^2}+{h^2\over\beta^2},
$$
we obtain
$$
Q\approx{\varsigma\over 2\pi\beta}  \exp\left\{- {h^2\over 2\beta^2} \right\}
\int_0^\infty {
\exp\left\{- { \left(s-h\varepsilon\right)^2\over2\varsigma^2}
\right\}
\over \varepsilon h+(1-\varepsilon)s
}
ds.
$$
By Laplace's method \cite{zorichII},
for sufficiently small $\varsigma>0\; (\varsigma\ll\varepsilon)$, 
the last integral is asymptotic to
$$\frac{\sqrt{2\pi}}{(h\varepsilon+(1-\varepsilon)\varepsilon h)\sqrt{1/\varsigma^2}}=\frac{\sqrt{2\pi}\varsigma}{h\varepsilon(2-\varepsilon)}.
$$
Hence,
$$Q\approx\frac\varsigma{2\pi\beta}\frac{\sqrt{2\pi}\varsigma}{h\varepsilon(2-\varepsilon)}\exp\left\{- {h^2\over 2\beta^2} \right\}
%=\frac{\varsigma^2}{\sqrt{2\pi}\beta{h}\varepsilon(2-\varepsilon)}
=\frac\beta{\sqrt{2\pi}h}\exp\left\{- {h^2\over 2\beta^2} \right\}.
$$
By the same reasoning the error term from (\ref{tail}) is of order
$$
\exp\left\{- {h^2\over 2\beta^2} \right\}\times O\left(\frac1\varepsilon\left(\frac\beta h\right)^3\right),
$$
which  gives
(\ref{5.3a}).
\hfill\qed

\subsection{The randomly perturbed map: random slope}
Consider a process
\begin{equation}\label{xn}Y_n=\mu(1+\sigma r_{1,n})Y_{n-1}+\sigma r_{2,n},\quad n\ge1,\end{equation}
where
 $(r_{1,n}, r_{2,n})_{n=1}^\infty$  are IID copies of a two dimensional  random vector $(r_1,r_2)$. 
Here, we assume that $(r_1,r_2)$ has bivariate normal distribution with mean vector $0$ and 
covariance matrix $\Sigma_2=[\sigma_{i,j}]$, where $\sigma_{i,j}=\cov(r_i,r_j)$, $1\le i,j\le 2$.
We  assume that the entries $\sigma_{i,j}$ are of order 1 in a sense that they do not depend on 
other parameters. 
Recall that the probability density function 
of a multivariate normal random vector $(r_1,\dots,r_d)$ 
with mean vector 0 and covariance matrix $\Sigma$ is given by
$$
\frac1{\sqrt{(2\pi)^d\mbox{det}(\Sigma)}}
\exp\left\{-\frac12x^T\Sigma^{-1}x\right\},\quad x=(x_1,\dots,x_d)^T.
$$%%\ee 
and we denote such vectors by  $N(0,\Sigma)$.

For a given $h>0$, let
$$\tau=\inf\{k\ge1:\ Y_k>h\}.$$

\begin{thm}\lbl{thm:randslope}
Suppose that $h$ and $\mu\in (0,1)$ are both of order 1 and $\sigma \ll1$ so that the following
condition holds
\begin{equation}\label{stabcond}\gamma:=\mu\E|1+\sigma r_1|<1.
\end{equation}
Then $\tau$ is asymptotically geometric 
RV with parameter
%\begin{equation}p\approx
%\frac\sigma{c\sqrt{2\pi}}e^{-\frac{c^2}{2\sigma^2}},\label{p-asympt}\end{equation}
\begin{equation}p=
\frac\sigma{c\sqrt{2\pi}}e^{-\frac{c^2}{2\sigma^2}}\Big(1+O(\sigma^2)\Big),\label{p-asympt}\end{equation}
where a positive constant $c$  depends on $h$, $\mu$, and $\Sigma_2$, but not on $\sigma$. 
\end{thm}

As before, we first establish convergence of $\{ Y_n\}$ and characterize the limit.
Iteration of (\ref{xn}) yields
\begin{eqnarray}\nonumber
Y_n&=&\mu(1+\sigma r_{1,n})Y_{n-1}+\sigma r_{2,n}=\mu(1+\sigma r_{1,n})\left(\mu(1+\sigma r_{1,n-1})Y_{n-2}+\sigma r_{2,n-1}\right)+
\sigma r_{2,n}\\&=&\dots=\mu^nY_0\prod_{j=1}^n(1+\sigma r_{1,j})+\sigma
\sum_{j=0}^{n-1}\mu^{j} r_{2,n-j}\prod_{\ell=n-j+1}^n(1+\sigma r_{1,\ell}),
\label{xin} \end{eqnarray}
where as usually, $\prod_{j=k}^m(\ *\ )=1$ if $k>m$.

\begin{lem}\lbl{lem:randslope-convergence}
\begin{equation}\label{defofxi}
Y_n\stackrel{d}{\longrightarrow} 
Y\stackrel{d}{=}\sigma\sum_{j=0}^\infty\mu^{j}g_{2,j}\prod_{\ell=0}^{j-1}(1+\sigma g_{1,\ell}), \; n\to\infty,
\end{equation}
where $(g_{1,j},g_{2,j}), \; j=0,1,2,\dots$ are IID copies of two-dimensional random vector, 
which is equal in distribution to  $(r_1,r_2)$.
\end{lem}
\pf (Lemma \ref{lem:randslope-convergence}):
First, we show that $Y$ is well-defined as the series in (\ref{defofxi}) converges almost surely. 
To see this, note that the summands
$$
g_{2,j}\prod_{\ell=0}^{j-1}(1+\sigma g_{1,\ell})$$
are martingale differences with respect to the natural filtration. 
By triangle inequality, independence, and
(\ref{stabcond}),
\begin{eqnarray*}
&&\E\left|\sigma\sum_{j=0}^m\mu^jg_{2,j}\prod_{\ell=0}^{j-1}(1+\sigma g_{1,\ell})\right|\le
\sigma\E|g_2|
\sum_{j=0}^m\mu^{j}\E\left|\prod_{\ell=0}^{j-1}(1+\sigma g_{1,\ell})\right|
\\&&\qquad=
\sigma\E|g_2|
\sum_{j=0}^m\mu^{j}\left(\E|1+\sigma r_{1}|\right)^j=
\frac{\sigma\E|g_2|}{1-\gamma}(1-\gamma^{(m+1)})
\le \frac{\sigma\E|g_2|}{1-\gamma}.
\end{eqnarray*}
Hence, the partial sums of the right--hand side of (\ref{defofxi}) form an $L_1$--bounded martingale which converges almost surely by the martingale convergence theorem (see e.g. \cite{steele}).
For every $n\ge1$
$$
\sigma
\sum_{j=0}^{n-1}\mu^{j} r_{2,n-j}\prod_{\ell=n-j+1}^n(1+\sigma r_{1,\ell})\stackrel{d}=\sigma
\sum_{j=0}^{n-1}\mu^{j} g_{2,j}\prod_{\ell=0}^{j-1}(1+\sigma g_{1,\ell}).
$$
Since the sequence on the right converges almost surely and the almost sure convergence implies convergence in distribution, we infer that the sequence on the left converges in distribution. To conclude that $Y_n\stackrel{d}\to Y$ it is enough to show that
 the first term  on the right--hand side of (\ref{xin}) converges to 0 in probability. But that is clear since we have
$$\E\left|Y_0\mu^n\prod_{j=1}^n(1+\sigma r_{1,j})\right|=
|Y_0|\mu^{n}\prod_{j=1}^n\E|1+\sigma r_{1,j}|=|Y_0|\gamma^{n}.
$$
Hence, by Markov inequality it goes to 0 in probability.
\hfill\qed

\pf (Theorem \ref{thm:randslope}):\;
The proof follows the lines of the proof of Theorem \ref{thm:addpert}.
The main complication in treating the present case is that we know  less  about 
the distribution of $Y_n$ than in before. Nonetheless, we will argue that for large $n$
\begin{equation}\lbl{condprob}
p_n:=\frac{\P(\tau=n)}{\P(\tau\ge n)}=\P(\mu(1+\sigma r_{1,n})Y_{n-1}+\sigma
r_{2,n}>h|Y_{n-1}\le h)\end{equation}
is approximately constant.
For this, we rewrite the right hand side of (\ref{condprob})  as
$$\frac{\P(\mu(1+\sigma r_{1,n})Y_{n-1}+\sigma
r_{2,n}>h,Y_{n-1}\le h)}{\P(Y_{n-1}\le h)}
,$$
and since the denominator converges to $\P(Y\le h)$  we focus on the numerator. 
Let  $(r_1,r_2)$ be a generic vector distributed like $(r_{1,n},r_{2,n})$
and independent  of $Y$. Since for every $n\ge1$, $(r_{1,n},r_{2,n})$ is independent of $Y_{n-1}$, as $n\to\infty$ we have
$$(r_{1,n},r_{2,n},Y_{n-1})\stackrel d\longrightarrow(r_1,r_2,Y).$$
Thus,
$$
\P(\mu(1+\sigma r_{1})Y_{n-1}+\sigma r_{2}>h,Y_{n-1}\le h)\longrightarrow
\P(\mu(1+\sigma r_{1})Y+\sigma r_{2}>h,Y\le h),
$$
which establishes  the existence of $p=\lim_{n\to\infty}p_n$.

To estimate $p$, we first recall that  $(r_1,r_2)$ is bivariate normal if and only if every linear combination of $r_1$ and $r_2$ %$\alpha_1r_1+\alpha_2r_2$ 
is a normal RV. Hence, conditionally on $Y=y$, 
$\sigma(\mu yr_1+r_2)$ is $N(0,\sigma^2\sigma_y^2)$ RV, where 
\be\lbl{sig_y}\sigma_y^2=\sigma_{22}^2+\mu^2y^2\sigma_{11}^2+2\mu y\sigma_{12}.\ee  
Therefore,
\begin{eqnarray*}&&\P(\mu(1+\sigma r_1)Y+\sigma r_2>h,\ Y\le h)=
\P(\sigma(\mu Y r_1+r_2)>h-\mu Y,Y\le h)
\\&&\qquad=
\int_{-\infty}^h\P(Z>\frac{h-\mu y}{\sigma\sigma_y})dF_Y(y)=\int_{-\infty}^h\left(1-\Phi\left(\frac{h-\mu y}{\sigma\sigma_y}\right)\right)dF_Y(y)\\
&&\qquad=\left(1-\Phi\left(\frac{h-\mu y_0}{\sigma\sigma_{y_0}}\right)\right)\P(Y\le h),
\end{eqnarray*}
where 
%%as above $Z$ stands for a RV with standard normal distribution and
 $-\infty<y_0<h$ by the mean value theorem. Hence,
$$
p=\frac{\P(\mu(1+\sigma r_1)Y+\sigma r_2>h,\ Y\le h)}{\P(Y\le h)}=
1-\Phi\left(\frac{h-\mu y_0}{\sigma\sigma_{y_0}
}\right).
$$
Let $c:=c(y_0)$ where
$$
c(x)=c_{h,\mu,\Sigma_2}(x):={h-\mu x\over \sigma_x}=\frac{h-\mu x}{\sqrt{\mu^2\sigma_{11}^2x^2+2\mu\sigma_{12}x+\sigma_{22}^2}}.
$$
Then, by (\ref{tail})
$$p=1-\Phi\left({c\over\sigma}\right)=
\frac\sigma{c\sqrt{2\pi}}e^{-\frac{c^2}{2\sigma^2}}\left(1+O\left(\frac{\sigma^2}{c^2}\right)\right).
$$
Furthermore,
by elementary analysis we see that: 
\begin{itemize}
\item
$c(x)$ is increasing 
 on $x\in(-\infty, x^*)$ and decreasing on $x\in(x^*,\infty)$, where $$x^*=-\frac{\sigma_{11}^2+h\sigma_{12}}{\mu(h\sigma_{22}^2+\sigma_{12})},$$
\item
$ c(-\infty)=\sigma_{11}^{-1},\; 
 c(h)={(1-\mu)h\over\left((\mu h\sigma_{11})^2+2\mu\sigma_{12}h+\sigma_{22}\right)^{1/2}}={(1-\mu)h\over\left((\mu h\sigma_{11}+\sigma_{22})^2-2\mu h(\sigma_{11}\sigma_{22}-\sigma_{12})\right)^{1/2}}$,
and $c(x^*)$ is given by a quite unwieldy expression that depends  on $h$ and $\Sigma_2$ but not on $\mu$.
\end{itemize}
In particular,  $c$ is bounded away from $0$ and $\infty$ provided $\mu$ and $h$ are positive
and $\mu<1$. This  proves (\ref{p-asympt}).
\hfill\qed

\subsection{A two-dimensional randomly perturbed map}

In this subsection we consider the following two dimensional model:
\begin{eqnarray}
\lbl{6.11}
\xi_{n+1} &=&\mu\xi_n\left(1+\sigma r_{1,n+1}\right)+\sigma r_{2,n+1},\\
\lbl{6.12}
\eta_{n+1} &=& \lambda \eta_n +\epsilon\sigma r_{3,n+1} +\epsilon a_2\xi_n.
\end{eqnarray}
where $(r_{1,n},r_{2,n},r_{3,n})$, $n\ge1$, is a sequence of IID copies of 
  $(r_{1},r_{2},r_3)$ which, as follows form a discussion at the beginning of Section~\ref{sec:exit}  is  
assumed to be a trivariate normal random vector
$N(0,\Sigma_3)$, with $\Sigma_3=[\sigma_{i,j}]$, $1\le i,j\le 3$, where $\sigma_{i,j}=\cov(r_i,r_j)$ do not depend on any  parameters in (\ref{6.1}) and (\ref{6.2}). 
For positive $h_1,h_2 =O(1)$, we define
$$
\tau_\xi=\inf_{k\ge 1} \{ \xi_k > h_1\},
\qquad
\tau_\eta=\inf_{k\ge 1} \{ \eta_k > h_2\}.
$$
We are interested in $\tau=\min\{\tau_\xi, \tau_\eta\}$. We know the distribution of $\tau_\xi$ from
Theorem~\ref{thm:randslope}. As we will show below, under the suitable conditions the distribution of $\tau$
is again asymptotically geometric. Moreover, if  $\epsilon>0$ is small then $\tau_\eta$ has practically no effect on the distribution of $\tau$.

In order to be more precise, let us define
\be\lbl{AnGn}
A_n=\left[\begin{array}{cc}\mu(1+\sigma r_{1,n})&0\\
\epsilon a_2&\lambda\end{array}\right],\quad
G_n=\left[\begin{array}{c}r_{2,n}\\\epsilon r_{3,n}\end{array}\right],\quad\mbox{and}\quad
\Theta_n=\left[\begin{array}{c}\xi_{n}\\\eta_{n}\end{array}\right].
\ee
Then,  (\ref{6.11}) and (\ref{6.12}) are described by
\be\lbl{rec} 
\Theta_{n+1}=A_{n+1}\Theta_n+\sigma G_{n+1},\quad n\ge1.
\ee

\begin{thm}\label{thm:2d-geom}
Let $\mu,\sigma,\epsilon\in (0,1)$ be such that 
$\mu$ is of order 1
 and  $\sigma \ll1$ so that
condition (\ref{stabcond}) holds. Assume  $\epsilon\ll1$and set $\la=1-\eps$. 
Suppose further
that $h_1$ and  $h_2$ are of order 1. Then $\tau$ is approximately geometric 
RV with parameter $p$ satisfying
\begin{equation}
p\approx\frac\sigma{c\sqrt{2\pi}}e^{-\frac{c^2}{2\sigma^2}},
\label{p-as2d}\end{equation}
and where  the constant $c$ depends on $h_1$, $\mu$, and $\Sigma_3$ but not on $\sigma$. 
\end{thm}
The following lemma shows that $\{\Theta_n\}$ converges in distribution and describes the limit.
\begin{lem}\lbl{lm:2d-convergence}
\be\lbl{Ydist}
\Theta_n\stackrel{d}{\longrightarrow} X\stackrel d=\sigma\sum_{k=1}^\infty\left(\prod_{j=1}^{k-1}A_j\right)G_k,\; 
n\to\infty,
\ee
where $A_n$ and $G_n, n=1,2,dots$ are defined in (\ref{AnGn}).
Furthermore, this random vector $X$  satisfies the distributional equation
\be\lbl{stocheq}
X\stackrel d=A X+\sigma G,
\ee
where
\be\lbl{gentheta}A=\left[\begin{array}{cc}\mu(1+\sigma r_{1})&0\\
\epsilon a_2&\lambda\end{array}\right]\quad\mbox{and}\quad G=\left[\begin{array}{c}r_{2}\\
\epsilon r_{3}\end{array}\right],\ee
$(r_1,r_2,r_3)$ is $N(0,\Sigma_3)$ be generic copies of $A_n$ and $G_n$, and,  $X$ 
on the right hand side of (\ref{stocheq})  is independent of  $(A,G)$.
\end{lem}
\pf (Lemma \ref{lm:2d-convergence}):
Note first that each of the  sequences  $(A_n)$ and $(G_n)$ consists of IID random elements.
Let 
$(r_1,r_2,r_3)$ is $N(0,\Sigma_3)$ be generic copies of $A_n$ and $G_n$. By iterating (\ref{rec}),
we obtain
$$\Theta_n=A_n(A_{n-1}\Theta_{n-2}+\sigma G_{n-1})+\sigma G_n=\dots=\left(\prod_{k=0}^{n-1}A_{n-k}\right)\Theta_0+\sigma\sum_{k=1}^n\left(\prod_{j=0}^{n-k-1}A_{n-j}\right)G_k,$$
where, as usually, the product is set to be $1$ if its index  range is empty.
We have
$$\prod_{k=0}^{n-1}A_{n-k}=\left[\begin{array}{cc}\mu^n\prod_{k=1}^n(1+\sigma r_{1,k})&0\\T_n&\lambda^n\end{array}\right],$$
where
$$T_n=\epsilon a_2
\sum_{j=1}^{n}\la^{n-j}\prod_{k=1}^{j-1}(\mu(1+\sigma r_{1,k})).$$
%Hence, 
Set $\delta=\max\{\la,\mu\E|1+\sigma r_1|\}$ and note that by (\ref{stabcond}) $\delta<1$.
By triangle inequality and independence of $r_{1,k}$'s
$$\E|T_n|\le\epsilon a_2\sum_{j=1}^n\la^{n-j}\E\left|\prod_{k=1}^{j-1}(\mu(1+\sigma r_{1,k}))\right|=\epsilon a_2\sum_{j=1}^n\la^{n-j}\left(\mu\E|1+\sigma r_1|\right)^{j-1}\le\epsilon a_2 n \delta^{n-1}.$$
Similarly,
$$\mu^n\E\left|
\prod_{k=1}^n(1+\sigma r_{1,k})\right|=\left(\mu\E|1+\sigma r_1|\right)^n\le\delta^n.$$ It follows  that both components of $\left(\prod_{k=0}^{n-1}A_{n-k}\right)\Theta_0$ converge to $0$ in probability and thus, this term is negligible.

Since the sequences $(A_n)$ and $(G_n)$ are IID, for every $n\ge1$ we have
$$\sum_{k=1}^n\left(\prod_{j=0}^{n-k-1}A_{n-j}\right)G_k\stackrel d=
\sum_{k=1}^n\left(\prod_{j=1}^{k-1}A_j\right)G_k.$$
By the same argument as above we verify that both components of the sequence of partial sums on the right hand side
are Cauchy in $L_1$. Hence, the components of the series
$$\sum_{k=1}^\infty\left(\prod_{j=1}^{k-1}A_j\right)G_k,$$
converge in probability (and thus, in distribution).
Therefore, the sequence  $(\Theta_n)$ defined by (\ref{rec})  converges in distribution to a 
random vector $X$ defined in (\ref{Ydist}).
Furthermore, $X$  satisfies the distributional equation (\ref{stocheq}).
\hfill
\qed

\pf (Theorem~\ref{thm:2d-geom}):
For $h=(h_1,h_2)$ set $B_h:=(-\infty,h_1]\times(-\infty,h_2]$. Then
$$\{\tau=n\}=\{\Theta_j\in B_{h},\ j<n,\ \Theta_n\notin  B_{h}\},$$
so that
\begin{eqnarray*}\P(\tau=n)&=&\P(\Theta_n\notin B_{h}|\Theta_j\in B_{h},\ j<n)\P(\Theta_j\in B_{h},\ j<n)
\\&=&\P(A_n\Theta_{n-1}+\sigma G_n\notin  B_{h}|\Theta_{n-1}\in  B_{h})\P(\tau\ge n).
\end{eqnarray*}
Since $\Theta_n$ converge in distribution to $X$ we have
\begin{eqnarray}\nonumber
&&p_n:=\P(A_n\Theta_{n-1}+\sigma G_n\notin  B_{h}|\Theta_{n-1}\in  B_{h})=
\frac{\P(A_n\Theta_{n-1}+\sigma G_n\notin  B_{h},\Theta_{n-1}\in  B_{h})}
{\P(\Theta_{n-1}\in  B_{h})}\\
&&\qquad\qquad\longrightarrow
p:=\frac{\P(AX+\sigma G\notin  B_{h},X\in  B_{h})}
{\P(X\in  B_{h})},\quad\mbox{as}\quad n\to\infty.
\lbl{limit}\end{eqnarray}
It follows from (\ref{Ydist}) that $X$ is symmetric, so since both $h_1$ and $h_2$ are positive the denominator is at least $1/2$ and does not affect the asymptotics.

To handle the numerator, using (\ref{gentheta}),  denoting the components of $X$ by $X_1$ and $X_2$, and using the notation adopted in (\ref{sig_y})  we see that it is equal to
\begin{eqnarray}
&&\P((\mu(1+\sigma r_1)X_1+\sigma r_2,\eps a_2X_1+\lambda X_2+\eps\sigma r_3)\notin B_h, (X_1,X_2)\in B_h)\nonumber\\&&\quad=
\P(\mu(1+\sigma r_1)X_1+\sigma r_2>h_1, (X_1,X_2)\in B_h)\nonumber\\&&\quad\quad+
\P(\eps a_2X_1+\lambda X_2+\eps\sigma r_3>h_2, (X_1,X_2)\in B_h)\nonumber\\&&\qquad\quad-
\P(\mu(1+\sigma r_1)X_1+\sigma r_2>h_1,\eps a_2X_1+\lambda X_2+\eps\sigma r_3>h_2, (X_1,X_2)\in B_h)\nonumber\\&&\quad=
\P\left(\frac{\mu X_1 r_1+r_2}{\sigma_{X_1}}>\frac{h_1-\mu X_1}{\sigma\sigma_{X_1}},(X_1,X_2)\in B_h\right)
\nonumber\\&&\qquad+
\P\left(r_3>\frac{h_2-\epsilon a_2 X_1-\lambda X_2}{\epsilon\sigma},(X_1,X_2)\in B_h\right)
\nonumber\\&&\qquad-
\P\left(\frac{\mu X_1 r_1+r_2}{\sigma_{X_1}}>\frac{h_1-\mu X_1}{\sigma\sigma_{X_1}},r_3>\frac{h_2-\epsilon a_2 X_1-\lambda X_2}{\epsilon\sigma},(X_1,X_2)\in B_h\right).
\lbl{3rd}\end{eqnarray}
Conditionally on $(X_1,X_2)=(x_1,x_2)$,
$$Z_1:=\frac{\mu x_1 r_1+r_2}{\sigma_{x_1}},\quad\mbox{and}\quad Z_2:=\frac{r_3}{\sigma_{33}}$$
are $N(0,1)$ RVs. 
 Hence by letting $F_X(x_1,x_2)$  denote
the distribution function of $(X_1,X_2)$, we see that   the first of the last three probabilities is
\be\lbl{1st}
\int_{-\infty}^{h_2}\int_{-\infty}^{h_1}\left(1-\Phi\left(\frac{h_1-\mu x_1}{\sigma\sigma_{x_1}}\right)\right)dF_X(x_1,x_2),
\ee
Likewise, for the second of these probabilities we get
\be\lbl{2nd}
\int_{-\infty}^{h_2}\int_{-\infty}^{h_1}\left(1-\Phi\left(\frac{h_2-\epsilon a_2x_1-\lambda x_2}{\epsilon\sigma\sigma_{33}}\right)\right)dF_X(x_1,x_2).\ee
 We now note that if $\epsilon$ is of a smaller order than all other parameters (except possibly $\sigma$) then (\ref{tail})
implies that  (\ref{2nd}) (and hence also  (\ref{3rd})) are negligible when compared to (\ref{1st}). To analyze the behavior of (\ref{1st}) as a function of its parameters note that by the mean value theorem the quantity in (\ref{1st}) is equal to
$$
\left(1-\Phi\left(\frac{h_1-\mu x_0}{\sigma\sigma_{x_0}}\right)\right)
\int_{-\infty}^{h_2}\int_{-\infty}^{h_1}
dF_X(x_1,x_2)=
\left(1-\Phi\left(\frac{h_1-\mu x_0}{\sigma\sigma_{x_0}}\right)\right)\P(X\in B_h),$$
for some $-\infty<x_0<h$. Substituting this into (\ref{limit}) (and neglecting the terms that depend on $\epsilon$) we see that
$$p=\frac{\P(AX+\sigma G\notin B_h, X\in B_h)}{\P(X\in B_h)}\sim
1-\Phi\left(\frac{h_1-\mu x_0}{\sigma\sigma_{x_0}}\right).$$
If both $0<\mu<1$ and $h_1$ are of order $1$ we are in the same situation as with (\ref{p-asympt}).
This shows (\ref{p-as2d}).
\hfill
\qed

\subsection{Diffusive escape}
The exit problems for the stochastic difference equations analyzed in the 
previous subsections all feature the geometric escape mechanism. In the
simplest case when the evolution is given by Equation (\ref{5.1}), 
the geometric distribution characterizes the statistics of the times of exit
of the trajectories of (\ref{5.1}) from a certain neighborhood of the attracting fixed point.
In this subsection, we study another important in applications statistical
regime associated with the exit problem for (\ref{5.1}), the diffusive
regime. The role of the diffusive regime in characterizing the statistics
of the exit times for the trajectories of (\ref{5.1}) is twofold. First,
the geometric distribution approximates the distribution of the exit times
only for sufficiently large times, i.e. for large $n$. In this subsection, 
we show that in the intermediate range of $n$, i.e. when $n$ is neither 
too large nor too small, $Y_n$'s are approximated by the sums of the IID
RVs and, therefore, the level exceedance times are distributed
as those for random walks. We refer to this situation as the diffusive regime.
Second, we recall that to justify the geometric distribution in the proof 
of Theorem \ref{thm:addpert}, we implicitly assumed that the rate of attraction of the
fixed point is stronger than the noise intensity. Specifically, 
it is easy to see from the proof of Theorem \ref{thm:addpert} that
$\varsigma$ is required to be $o(\epsilon)$, $\epsilon=1-\lambda$.
The analysis in this subsection does not use this assumption. We show that
when the noise is stronger than the attraction of the fixed point (albeit both
are sufficiently small), the mechanism of escape of the trajectories from the 
basin of attraction of the fixed point changes from the geometric to diffusive.
Therefore, we conclude this section by pointing out to some features intrinsic 
to the diffusive escape. Specifically, we consider (\ref{5.1}), for which as
before, we define
\be\lbl{d.1}
\tau=\inf\{k\ge 1:\;Y_k>h\},
\ee
for given $h>0$. 
In contrast to the case considered in Section 3.2, here we assume
\be\lbl{d.2}
\varepsilon=O(\varsigma^\alpha), \;\;\alpha>0.
\ee
In Theorem \ref{eps=0} below, we show that in the present situation in the intermediate 
range of $n$, $Y_n's$ behave as sums of IID normal RVs. The behavior of the latter is 
well-known (cf, Lemma \ref{known}). 
\begin{figure}
\begin{center}
\epsfig{figure=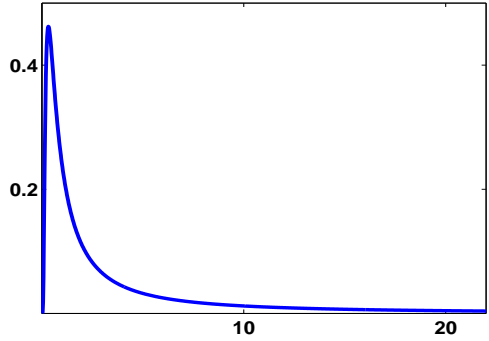, height=2.0in, width=3.0in, angle=0}
\end{center}
\caption{Probability density function corresponding to the distribution
$\Psi_a(y), a=1$. With a suitable $a>0$, $\Psi_a(y)$ approximates 
the distribution of the exit times in the diffusive escape.}
\label{f.3}
\end{figure}

Recall that $\Phi(x)$ stands for the
distribution function of an $N(0,1)$ RV and denote
\be\lbl{Psi} 
\Psi_a(x)=2\left(1-\Phi\left({a\over \sqrt x}\right)\right),\;\; a>0.
\ee
Note that $\Psi_a(x)$ is a probability distribution function on $\Re^+$
(see Fig. \ref{f.3}). 
\begin{thm}\lbl{eps=0}
Let the evolution of $Y_n, \;n=0,1,2,\dots$ be given by (\ref{5.1}). 
Suppose that  $\lambda=1-\varepsilon$ with 
$\varepsilon=O\left(\varsigma^{\alpha}\right),\;\alpha>0$.
Then for arbitrary positive $\beta_1$ and $\beta_2$ such that
$\beta_1+\beta_2<2\alpha/3$, for sufficiently small $\varsigma>0$,
\begin{equation}\lbl{distr_eps_0}
\P(\tau\le n)= \Psi_a(n)\Big(1+o(1)\Big), \quad a={h\over\varsigma},
\end{equation}
in the range $\varsigma^{-\beta_1}\ll n\ll\varsigma^{\frac{-2\alpha}{3}+\beta_2}.$
\end{thm}
\begin{rem}\lbl{range}
Since $\beta_{1,2}>0$ are arbitrary, $\Psi_a(n)$ practically approximates  $\P(\tau\le n)$
in the range $1\ll n\ll \varepsilon^{-2/3}$.
\end{rem} 
We will need the following auxiliary lemma~\cite[Theorem~2.2, Chapter III]{doob}.
It may be viewed as a quantified version of a reflection principle for random walk 
(see, e.g., \cite[Sec. 5.3, 5.4]{steele}). 
\begin{lem}\lbl{known}
Let $X_1,X_2,\dots$ be a sequence of independent, symmetric RVs and set 
$$S_k=\sum_{j=1}^kX_j,\quad\mbox{and}\quad S_k^*=\max_{1\le j\le k}S_j,\quad j\ge1.$$
Then for any $t,u>0$ the following
inequalities hold:
\begin{equation}\label{levy}
2\P(S_n\ge t+2u)-2\sum_{k=1}^n\P(X_k\ge u)\le\P(S_n^*\ge t)\le2\P(S_n\ge t).
\end{equation}
\end{lem}
\begin{rem}\lbl{rm:kw}
As was noticed by S.~Kwapie\'n a bit stronger   
version of the first inequality in (\ref{levy}) follows from a slight modification of 
the proof of Proposition~1.3.1 in \cite{kw}.
\end{rem}
\pf (Theorem \ref{eps=0}):  Without loss of generality, we assume that $Y_0=0$ 
(otherwise, apply the same argument to $Y_k-Y_0$). 
Note that the distributions of $\tau$ and $Y_k^*$ are linked by 
the following relation
$$
\P(\tau\le n)=\P(Y_n^*\ge h).
$$
Unwinding (\ref{5.1}) and using $Y_0=0$ gives
$$Y_k=\varsigma(\la^{k-1}r_1+\la^{k-2}r_2+\dots+\la r_{k-1}+r_k),
$$
which we write as $S_k+W_k$, where
\be\lbl{SW}
S_k:=\varsigma\sum_{j=1}^kr_j,\qquad W_k:=\varsigma\sum_{j=1}^{k-1}r_{j}(\la^{k-j}-1).
\ee
We will first show that the main contribution to $Y_n^*$ is from the $S_n^*$.
First, by subadditivity of  maxima,   
for any $0<h_1<h$, 
\begin{eqnarray}\nonumber
\P(Y_n^*\ge h)\le \P(S_n^*+W_n^*\ge h)
&\le&\P(S_n^*\ge h-h_1)+\P(W_n^*\ge h_1)\\
\label{*ubdd}
&\le&\P(S_n^*\le h-h_1)+\P\left(\left|W_n\right|^*\ge h_1\right).
\end{eqnarray}
Further, $Y_k \ge S_k-|W_k|$ so that
$$\P(S_n^*\ge h+h_1)\le \P(S_n^*\ge h+h_1, |W_n|^*<h_1)+
\P(|W_n|^*\ge h_1)\le \P(Y_n^*\ge h)+\P(|W_n|^*\ge h_1),$$
which, when combined with (\ref{*ubdd}) means that
\begin{equation}\lbl{*as}
\P(S_n^*\ge h+h_1)-\P(|W_n|^*\ge h_1)\le\P(Y_n^*\ge h)\le\P(S_n^*\ge h-h_1)+\P(|W_n|^*\ge h_1).
\end{equation}
First, we estimate $\P(|W_n|^*\ge h_1)$ in (\ref{*as}). To this end, we use
$1-\la^j=1-(1-\varepsilon)^j\le j\varepsilon$ to obtain
$$
\var(W_n)=\varsigma^2\sum_{j=1}^{n-1}(1-\la^j)^2\le\varsigma^2\varepsilon^2\frac{n^3}3=\varsigma^2n
\frac{\varepsilon^2n^2}3.
$$
Consequently, by (\ref{levy}) and (\ref{SW}), we have
\begin{eqnarray*}\P(|W_n|^*\ge h_1)&\le& 2\P(|W_n|\ge h_1)\le4\P(W_n\ge h_1)
=4\P\left(Z\ge\frac{ h_1}{\sqrt{\var(W_n)}}\right)\\
&\le&4\P\left(Z\ge\frac{ h_1}{\varsigma\sqrt n}\cdot\frac{\sqrt3}
{\varepsilon n}\right).
\end{eqnarray*}
Next, we turn to estimating the probabilities involving $S_n^*$ in (\ref{*as}).
By the second inequality in (\ref{levy}), for every $u>0$, we have
 \begin{equation}\label{ubdd}
 \P(S_n^*\ge h-h_1)\le2\P(S_n\ge h-h_1)=
  2\P\left(Z\ge \frac{h-h_1}{\varsigma\sqrt n}\right),
 \end{equation}
while the first one yields
\begin{eqnarray}
\P(S_n^*\ge h+h_1)&\ge&  2\P\left(S_n\ge h+h_1+2u\right)
-2\sum_{k=1}^n\P(\varsigma r_k\ge u)
\nonumber\\
&=&
2\P\left(Z\ge \frac{h+h_1+2u}{\varsigma\sqrt n}\right)-2n\P\left(Z\ge \frac u{\varsigma}\right).\lbl{lbdd}
\end{eqnarray}
The combination of (\ref{*as}), (\ref{ubdd}), and (\ref{lbdd}) yields
\begin{eqnarray}\lbl{combine_a}
\P(Y_n^*\ge h)&\ge&
2\P\left(Z\ge \frac{h+h_1+2u}{\varsigma\sqrt n}\right)-2n\P\left(Z\ge \frac u{\varsigma}\right)
-4\P\left(Z\ge\frac{ h_1}{\varsigma\sqrt n}\cdot\frac{\sqrt3}{\varepsilon n}\right),\\
\lbl{combine_b}
\P(Y_n^*\ge h)
&\le&
 2\P\left(Z\ge \frac{h-h_1}{\varsigma\sqrt n}\right)+
4\P\left(Z\ge\frac{ h_1}{\varsigma\sqrt n}\cdot\frac{\sqrt3}{\varepsilon n}\right)
\end{eqnarray}
To complete the proof, we need to chose $h_1$ and $u$ such that 
\be\lbl{desired}
{h_1\over \varsigma\sqrt{n}}=o(1),\quad {u\over\varsigma\sqrt{n}}=o(1),\quad {\varsigma\over u}=o(1),
\quad\mbox{and}\quad h_1^{-1}\varsigma\varepsilon n^{3/2}=o(1).
\ee
It is straightforward to verify that relations in (\ref{desired}) hold with
$h_1=\varsigma^{1+{3\beta\over_2}}$ and $u=\varsigma^{1-{\beta_1\over 2}},$ $\beta_{1,2}>0,$
$\beta_1+\beta_2<2\alpha/3$, and $n$ as in (\ref{distr_eps_0}).
\qed

\section{The Poincare map}
\setcounter{equation}{0}

In the present section, we consider the type I model, i.e. the randomly perturbed system with the
stochastic forcing acting via the fast subsystem (see (\ref{0.1}) and (\ref{0.2})). In the active
phase of bursting (when the system undergoes spiking), the trajectory of the randomly perturbed system
remains in the vicinity of the cylinder foliated by the periodic orbits of the fast subsystems,
(see Fig. \ref{f.4}a). The time that the trajectory spends near $L$ determines the duration of the
active phase. The goal of this section is to describe the slow dynamics near $L$. 
In particular, we will
estimate the distribution of the number of spikes in one burst.
To this end, we introduce a transverse to $L$ crossection $\Sigma$ (see Fig. \ref{f.4}a) and construct the 
first return map. Specifically, we estimate the change
in the state of the system after one cycle of rotation of the trajectory around $L$.
The construction of the first return map for (\ref{0.1}) and (\ref{0.2}) is done in analogy to that for
the deterministic models of bursting (see \cite{medvedev05, lee_terman}). However, the treatment of
the randomly perturbed system requires certain modifications. First, we have to resolve the ambiguity in
the notion of the first return time. The latter is due to the fact that generically a trajectory of
the randomly perturbed system makes multiple crossings with $\Sigma$ during each cycle around $L$.
We refer the reader to the comments following Theorem 2.3 in \cite{FW} for an explicit example
illustrating this effect.
For the randomly perturbed system, we define the time of the first return so that it approaches the
first-return time of the underlying deterministic system in the limit of vanishing random perturbation.
The definition of the first return time motivates the definition of the Poincare map (see Definition 4.1).
In Sections 4.1 and 4.2, we use asymptotic expansions to construct the linear approximation
for the Poincare map of the fast subsystem. Here, we use an obvious observation that on finite
time intervals and for sufficiently small $\epsilon>0$, the slow variable typically remains in
an $O(\epsilon)$ neighborhood of its initial value. Therefore, for finite times the Poincare map
of the fast subsystem captures the dynamics of the full system.
Since we are interested in long term behavior of the system, to complete the description of the first return
map we also need to track the (small) changes in the slow variable after each cycle of oscillations.
This is done in Section 4.3, where we derive a $1D$ map for the slow variable. The combination of
the $1D$ Poincare map for the fast subsystem and that for the slow variable provides the first return
map for the full problem (\ref{0.1}) and (\ref{0.2}). The linear approximation
of the $2D$ map is used in Section 4.4 to estimate the distribution of the number of spikes in one burst
for the type I model.
Effectively, the problem is reduced to the exit problem for a $1D$ linear randomly perturbed map.
For the latter problem, we have already developed necessary analytical  tools in Section 3. 
Finally, in Section 4.5, we comment on the straightforward modifications necessary to extend the 
analysis of this section to cover type II models.

\subsection{Preliminary transformations}

\begin{figure}
\begin{center}
{\bf a}\epsfig{figure=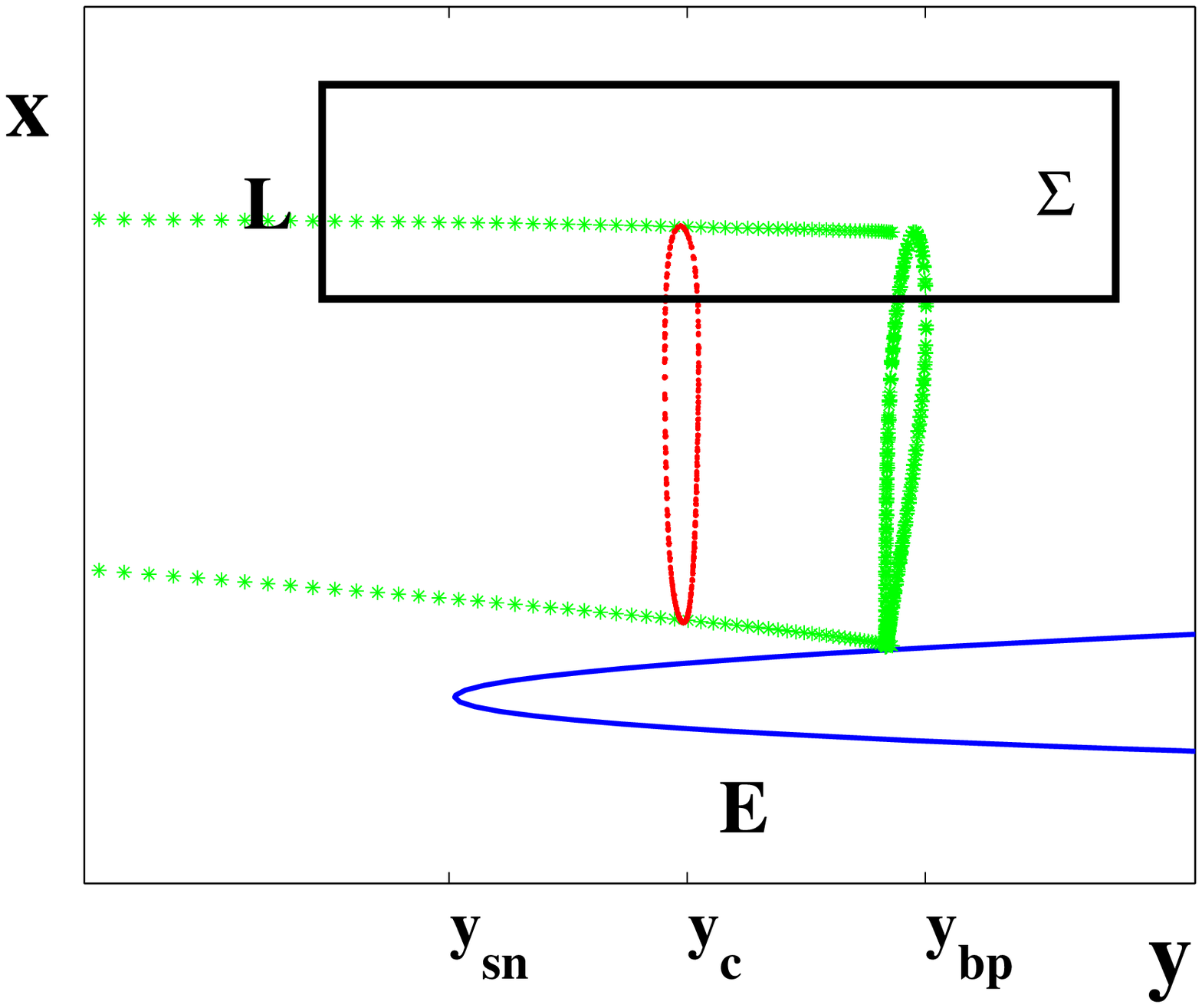, height=3.0in, width=4in, angle=0}\quad
{\bf b}\epsfig{figure=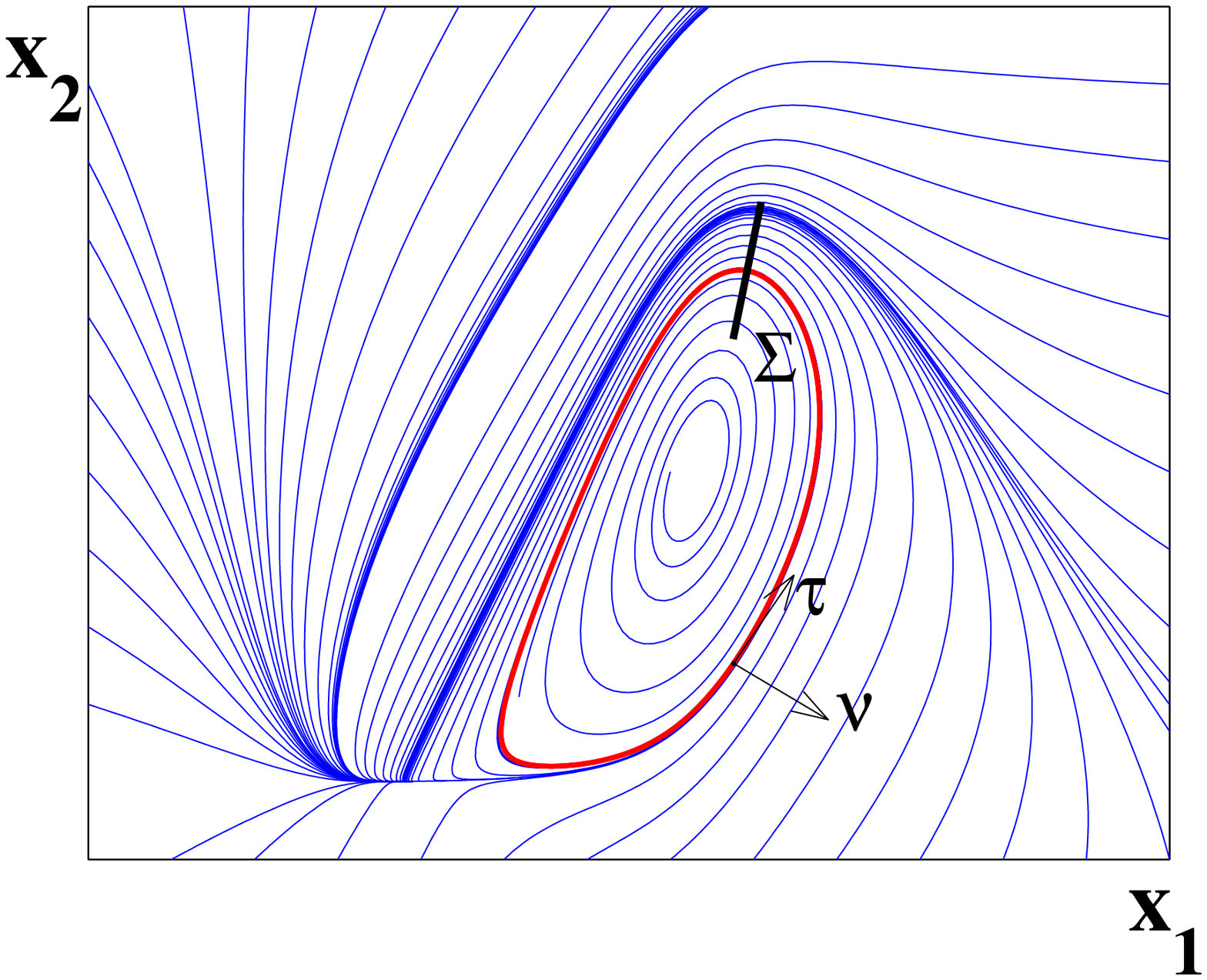, height=3.0in, width=4in, angle=0}
\end{center}
\caption{(a) Crossection $\Sigma$ is used in the construction of the first return map.
(b) The phase plane of the fast subsystem (\ref{1.3}) for $y\in (y_{sn}, y_{bp})$.
}
\label{f.4}
\end{figure}

Recall that $\Sigma$ stands for the  transverse section located as shown schematically
in Fig. \ref{f.4}a.
Let $y_0< y_{bp}$ be outside an $O(\sigma)$ neighborhood of $y_{bp}$,
and $x_0=\left(x_0^1, x_0^2\right)^T\in \Sigma$ be from an $O(\sigma)$ neighborhood of $L$.
Consider an initial value problem for (\ref{0.1}) and (\ref{0.2}) with initial data
$\left(x_0, y_0\right)$. By standard results from the asymptotic theory for randomly perturbed
systems \cite{FW}, we have the following estimate
\be\lbl{2.1}
y_t=y_0+O(\epsilon),
\ee
valid on a finite interval of time $t\in [0, \bar t]$. Here and below, for a small parameter
$\epsilon>0$, the symbols $O(\epsilon)$ 
and $o(\epsilon)$ in the asymptotic expansions of the random functions mean that the corresponding
relations hold almost surely (a.s.). Specifically,  
$\psi_t(\epsilon)=O(\epsilon)$ for $t\in [t_1, t_2]$ means that there exists 
$\epsilon_0>0$ such that 
$$ 
\sup_{\tiny
\begin{array}{c}
t\in [t_1, t_2]\\
\epsilon\in[0, \epsilon_0]
\end{array}
} 
\left|\epsilon^{-1}\psi_t(\epsilon)\right|<\infty \quad \mbox{a.s.}.
$$ 
In a similar fashion, we interpret $\psi_t(\epsilon)=o(\epsilon)$ when
$\psi_t(\epsilon)$ is a random function.  

By plugging in (\ref{2.1}) into (\ref{0.1}),
we obtain the following SODE
\be\lbl{2.2}
dx_t = f\left(x_t\right)dt+\sigma p(x_t)dw_t +O(\epsilon),
\ee
where $f\left(x\right):=f\left( x,y_0\right)$, $p(x):=p\left(x, y_0\right),$
and $y_0$ is fixed.
Equation (\ref{2.2}) with $\epsilon=\sigma=0$ has an exponentially  orbitally stable periodic
solution $x=\phi(t, y_0)$ of period $\Tb(y_0)$:
$$
L(y_0)=\left\{ x=\phi(\theta, y_0):\; \theta\in \left[0, \Tb(y_0) \right)\right\}
\qquad (\mbox{cf}. \;(\ref{1.4})).
$$
 To simplify the notation, throughout the analysis
of the fast subsystem, we will omit to indicate the dependence on $y_0$ when refer
to $L$, $\phi$, and $\Tb$.
At each point $x=\phi(\theta)\in L$, we define vectors
\be\lbl{2.3}
\tau(\theta)=\left( f^1(x), f^2(x)\right)^T\quad\mbox{and}\quad \nu(\theta)=J f(x),\;\;
\mbox{where}\;\; J=\left(
\begin{array}{cc}
0& -1\\ 1 &0 \end{array}\right),
\ee
pointing in the tangential and normal directions, respectively.
To study the trajectories of (\ref{2.2}) in a small neighborhood of $L$,
it is convenient to rewrite (\ref{2.2}) in normal coordinates $\left(\theta, \xi\right)$ \cite{hale}:
\be\lbl{2.4}
x=\phi(\theta)+\xi\nu(\theta),\quad \theta\in [0,\Tb).
\ee

\begin{lem}
For sufficiently small $\delta>0$ Equation (\ref{2.4}) defines a smooth change of coordinates in
\be\lbl{2.4a}
B_\delta=\{x=\phi(\theta)+\xi\nu(\theta):\; \left|\xi\right|<\delta,\; \theta\in [0,\Tb)\}.
\ee
In new coordinates, (\ref{2.2}) has the following form:
\begin{eqnarray}\lbl{2.5}
 d\theta_t &=& (1 + b_1(\theta_t)\xi_t)dt +\sigma h_1(\theta_t, \xi_t)
\left( 1+b_2(\theta_t)\xi_t\right) d w_t +O(\eps, \delta^2, \sigma^2),\\
\lbl{2.6}
d\xi_t &=& a(\theta_t)\xi_tdt+ \sigma h_2(\theta_t,\xi_t) dw_t+O(\eps, \delta^2, \sigma^2),
\end{eqnarray}
where  smooth functions $a(\theta), b_1(\theta), $ and $b_2(\theta)$ are $\Tb-$periodic and
\be\lbl{2.7}
0<\mu:=\exp\left(\int_0^\Tb a(\theta)d\theta\right) =\exp\left(\int_0^\Tb \mathrm{div} f\left(\phi(\theta)\right)\right)<1,
\ee
\be\lbl{2.8}
h_1(\theta,\xi)={<p,\tau>\over <\tau,\tau>}= {p^1f^1+p^2f^2\over \left|f\right|^2},\;\;
h_2(\theta,\xi)={<p,\nu>\over <\tau,\tau>}= {p^2f^1-p^1f^2\over \left|f\right|^2}.
\ee
\end{lem}

\pf:\;
The proof of the lemma follows the lines of the proof of Theorem VI.1.2 in \cite{hale}.
Let $z=(z^1,z^2)^T:=(\theta, \xi)^T$ and denote the transformation in (\ref{2.4}) by
\be\lbl{2.8a}
x=v(z),\; z\in B_\delta.
\ee
Note
$$
\left|Dv(\theta,0)\right|=\left|\begin{array}{cc}
{\phi^1}^\prime (\theta)& -f^2\left(\phi(\theta)\right)\\
{\phi^2}^\prime (\theta)&  f^1\left(\phi(\theta)\right)\\
\end{array}
\right|=\left|f\left(\phi(\theta)\right)\right|^2\ne 0,\; \theta\in[0,\Tb).
$$
Therefore, for sufficiently small $\delta>0$,  (\ref{2.8a}) defines a smooth invertible transformation
in $B_\delta$.
Denote the inverse of $v$ by $z=u(x),\; x\in v(B_\delta)$
and note that
\be\lbl{2.8b}
\left[Du(x)\right]^{-1}=Dv(z),\; x\in v(B_\delta).
\ee
By It\^{o}'s formula, we have
\be\lbl{2.8c}
dz_t=Du(x_t)dx_t + O(\sigma^2)dt
\ee
and, therefore,
\be\lbl{2.8d}
Dv(z_t)dz_t= dx_t + O(\sigma^2)dt.
\ee
By recalling that $z=(\theta,\xi)$ and after plugging in (\ref{2.2})
into (\ref{2.8d}), we obtain
\begin{eqnarray}\nonumber
\left[ {d\phi(\theta_t) \over d\theta}+ {d \nu(\theta_t)\over d\theta} \xi_t\right]
d\theta_t +\nu(\theta_t) d\xi_t& = &\left(f\left(\phi(\theta_t)\right)+
Df\left(\phi(\theta_t)\right)\nu(\theta_t)\xi_t
      +Q(\theta_t, \xi_t)\right)dt \\ \lbl{2.9}
&+&
\sigma p  dw_t +O(\epsilon,\sigma^2),
\end{eqnarray}
where
$$
Q(\theta,\xi)=f\left(\phi(\theta)+\xi\nu(\theta)\right)-f\left(\phi(\theta)\right)-Df\left(\phi(\theta)\right)\nu(\theta)\xi
=O\left(\xi^2\right),\quad \left|\xi\right|<\delta.
$$
Note that
\begin{eqnarray}\lbl{2.10}
{d\phi(\theta)\over d\theta}&=&f\left(\phi(\theta)\right)=\tau(\theta),
\quad \tau^T(\theta)\tau(\theta) =\nu(\theta)^T\nu(\theta)=\left|f\left(\phi(\theta)\right)\right|^2,\\
\lbl{2.11}
{d\nu (\theta)\over d\theta}&=& {d\over d\theta} Jf\left(\phi(\theta)\right)=
JDf\left(\phi(\theta)\right) f\left(\phi(\theta)\right).
\end{eqnarray}
Taking into account (\ref{2.10}) and (\ref{2.11}), we project (\ref{2.9})
onto the subspace spanned by $\tau(\theta_t)$ and after some algebra obtain:
\be\lbl{2.11a}
\dot \theta_t = 1 + {f^T Q +f^T\left[ DfJ -J Df\right]f\xi_t +\sigma f^T p\dot w_t +O(\epsilon)
\over f^Tf +f^TJDf f\xi_t}.
\ee
Here and for the rest of the proof, for brevity we use the following notation:
$$
f:=f\left(\phi(\theta_t)\right),\quad Q:=Q(\theta_t, \xi_t),\quad\mbox{and}\quad \nu:=\nu(\theta_t).
$$
Equation (\ref{2.11a}) can be rewritten as (\ref{2.5}) with
\begin{eqnarray*}
b_1(\theta_t) &=& {1\over \left|f\right|^2}f^T\left[ DfJ -J Df\right]f,\\
b_2(\theta_t) &=& {1\over \left|f\right|^2}f^T JDf f.
\end{eqnarray*}
Similarly, by projecting (\ref{2.9}) onto the subspace spanned by $\nu(\theta)$ and
using (\ref{2.10}) and (\ref{2.9}), we derive
$$
\dot \xi_t = a(\theta_t)\xi_t+\sigma h_2(\theta_t) \dot w_t +O\left(\delta^2\right),
$$
where
$$
a(\theta_t)={1\over\left|\nu\right|^2} \nu^T\left[ Df \nu +{d\nu\over d\theta}\right]
-{2\nu^T\over\left|\nu\right|^2} {d\nu\over d\theta}.
$$
The expression in the square brackets can be simplified as follows:
$$
\nu^T\left[ Df \nu +{d\nu\over d\theta}\right]=f^T \left[ J^T Df J+Df \right]=
\mbox{div} f\left(\phi(\theta)\right) \left|f\right|^2.
$$
Also,
$$
{2\nu^T\over\left|\nu\right|^2} {d\nu\over d\theta}={2\over\left| f \right|^2} f^T J^T{d\over d\theta} J f =
{2\over\left| f \right|^2} f^T {d\over d\theta} f ={1\over\left| f \right|^2}{d\over d\theta}\left| f \right|^2
={d\over d\theta}\ln\left|f\left(\phi(\theta)\right)\right|^2.
$$
Therefore,
\be\lbl{2.12}
a(\theta)=\mbox{div} f\left(\phi(\theta)\right)-
{d\over d\theta}\ln\left|f\left(\phi(\theta)\right)\right|^2.
\ee
Equation (\ref{2.12}) implies (\ref{2.7}), since the integral over $[0,\Tb]$ of the last
term on the right hand side of (\ref{2.12}) is zero.
\qed

\subsection{The Poincare map for the fast subsystem}
In the present subsection, we analyze the trajectories of the
randomly perturbed system (\ref{2.2}) lying close
to the limit cycle $L(y_0),\; y_0<y_{bp}$. To this end, we consider an
IVP for (\ref{2.5}) and (\ref{2.6}) subject to the initial
condition:
\be\lbl{9.1}
\theta_0=0\quad\mbox{and}\quad\left|\xi_0\right|<\delta.
\ee
Throughout this section, we assume (even when it is not stated explicitly)
that $\delta>0$ is sufficiently small. 
It will be convenient to view 
the range of $\theta_t$ as $\mathbb{R}^1$ rather than a circle. Equation (\ref{2.4}) 
provides the transformation of $(\theta_t, \xi_t)$ to the Cartesian 
coordinates even when $\theta_t$ exceeds $\mathcal{T}$. 

We now turn to the construction of the Poincare map. Condition $\theta=0$ defines a 
transverse crossection of $L(y_0)$, $\Sigma$. The trajectory of the deterministic system 
(\ref{2.5}) and (\ref{2.6}) with $\sigma=0$ returns to $\Sigma$ in time
$\Tb+O(\xi_0)$. To define the Poincare map for the randomly perturbed system, we also use
another  transverse crossection  $\tilde\Sigma$, which is
located at an $O(1)$ distance away from $\Sigma$. Let $(\theta_t,\xi_t)$ be 
the solution of the IVP (\ref{2.5}), (\ref{2.6}), and (\ref{9.1}) and 
$$
\tilde T=\inf\{t>0:\; (\theta_t,\xi_t)\in \tilde\Sigma\}.
$$
\begin{df}\lbl{return_time} By the time of the first return of the trajectory 
(\ref{2.5}), (\ref{2.6}), and (\ref{9.1})
to $\Sigma$, we call stopping time $T$ such that 
\be\lbl{stop}
T=\inf\{t>\tilde T:\; \theta_t=\mathcal{T}\}.
\ee 
The first return map for (\ref{2.5}), (\ref{2.6}), and (\ref{9.1}) is defined 
as
$$
\bar\xi=P(\xi_0),\quad\mbox{where}\quad\bar\xi=\xi_T.
$$
\end{df}
In the remainder of this subsection, we compute the linear part of the Poincare map.
In the asymptotic expansions below, we omit to indicate the dependence
of the remainder terms on $\epsilon>0$. The latter is assumed to be sufficiently
small so that it has no effect on the leading order approximation of the Poincare map.

The following notation is reserved for four functions, which will appear 
frequently in the asymptotic expansions below:
$$
\begin{array}{ll}
A(t,s)=\exp\{\int_s^t a(u)du\},& A(t)=A(t,0),\\
B(t,s)=\int_s^t A(u,s)b_1(u)du, & B(t)=B(t,0).
\end{array}
$$
\begin{lem}
On a finite time interval $t\in[0,\bar t], 0<\bar t<\infty,$ the solution of the 
IVP (\ref{2.5}), (\ref{2.6}) and (\ref{9.1}) admits the following
asymptotic expansion
\begin{eqnarray}
\dot\het &=& \theta_t^{(0)}+\sigma\t1+O(\sigma^2,\xi_0^2), \lbl{9.4}\\
\dot\xt  &=& \xi^{(0)}_t+\sigma\x1+O(\sigma^2,\xi_0^2).\lbl{9.5}
\end{eqnarray}
The leading order coefficients are given by
\begin{eqnarray}\lbl{9.6}
\theta_t^{(0)}&=& t+\xi_0B(t)+O(\xi_0^2),\\
\xi^{(0)}_t&=&\xi_0A(t)+O(\xi_0^2). \lbl{9.7}
\end{eqnarray}
The first order terms are given by Gaussian diffusion 
process $z_t=\left(\theta^{(1)}_t,\x1\right)^T$:
\be\lbl{9.8}
z_t=\int_0^t U(t,s)h(s)dw_s+O(\xi_0),
\ee
where
\be\lbl{9.9}
U(t,s)=\left(\begin{array}{cc}
               1 & B(t,s)\\
               0 & A(t,s)
\end{array}
\right),\quad
h(t):=h(t,0)=\left(h_1(t,0),h_2(t,0)\right)^T.
\ee
\end{lem}
\pf:\; The procedure for constructing asymptotic expansions of solutions
for a class of IVP, which includes (\ref{2.5}), (\ref{2.6}) and (\ref{9.1})
can be found in \cite{BL62, FW}. These sources also contain the estimates
controlling the remainder terms. The coefficients $\het^{(0,1)}$ and $\xt^{(0,1)}$
are determined as follows. By plugging in (\ref{9.4}) and (\ref{9.5}) into
(\ref{2.5}) and (\ref{2.6}) and extracting the coefficients multiplying 
different powers of $\sigma$, one obtains IVPs for the functions on the 
right hand sides of (\ref{9.4}) and (\ref{9.5}). Specifically, for the
leading order terms we have the following IVP:
\begin{eqnarray}
\dot\theta_t^{(0)}&=&1+b_1\left(\theta_t^{(0)}\right)\xi^{(0)}_t,\lbl{9.10}\\
\dot\xi^{(0)}_t&=& a\left(\theta_t^{(0)}\right)\xi^{(0)}_t,\lbl{9.11}\\
\xi_0^{(0)}&=&\xi_0,\; \theta_t^{(0)}=0. \lbl{9.12}
\end{eqnarray} 
To the next order,
\begin{eqnarray}
\dot z_t&=&\Lambda(t,\xi_0)z_t + h\left(\theta_t^{(0)},\xi^{(0)}_t\right)dw_s\lbl{9.13},\\
z_0&=&0, \lbl{9.14}
\end{eqnarray}
where $z_t=\left(\t1,\x1\right)^T,\; h=(h_1,h_2)^T,$ and
\be\lbl{9.15}
\Lambda(t,\xi_0)=\left(
\begin{array}{cc}
b_1^\prime\left(\theta_t^{(0)}(\xi_0)\right)\xi^{(0)}_t(\xi_0) & b_1\left(\theta_t^{(0)}(\xi_0)\right) \\
b_1\left(\theta_t^{(0)}(\xi_0)\right)\xi^{(0)}_t(\xi_0) & a\left(\theta_t^{(0)}(\xi_0)\right) \\
\end{array}
\right).
\ee
Here, we explicitly indicated the dependence of the leading order coefficients
on $\xi_0$ and used prime to denote the differentiation with respect to $\theta$.
Formulae (\ref{9.6})-(\ref{9.9}) in the statement of the lemma
follow from (\ref{9.10})-(\ref{9.15}). The details can be found in the appendix to this 
paper.\\
\qed 
Next, we calculate the time of the first return.
\begin{lem}
The time of the first return is given by 
\be\lbl{9.16}
T=T^{(0)}+\sigma T^{(1)} +o(\sigma)+O(\xi_0^2),
\ee
where
\begin{eqnarray}\lbl{9.17}
T^{(0)}&=&\Tb-\xi_0B(\Tb)+O(\xi_0^2),\\
T^{(1)}&=&-\sigma\theta_\Tb^{(1)}=
-\sigma\int_0^\Tb \left[ h_1(u)+B(\Tb,u)h_2(u)\right]dw_u.  \lbl{9.18}
\end{eqnarray}
\end{lem}
\pf:\; 
From the definition of the first return time, (\ref{9.4}), and (\ref{9.6}), we have
\be\lbl{9.19}
T+\xi_0B(T)+\sigma\theta_T^{(1)} + O(\sigma^2,\xi_0^2)=\Tb\; \mbox{a.s.}.
\ee
Thus,
\be\lbl{9.20}
\lim_{\sigma\to 0} T=T^{(0)}(\xi_0)\; \mbox{a.s.},
\ee
where $T^{(0)}(\xi_0)$ is found from the following equation 
\be\lbl{9.21}
T^{(0)}(\xi_0)+\xi_0B\left(T^{(0)}(\xi_0)\right)+O(\xi_0^2)=\Tb.
\ee
Equation (\ref{9.21}) implies (\ref{9.17}). Furthermore, the combination
of (\ref{9.17}), (\ref{9.19}), and (\ref{9.20}) yields (\ref{9.18}).\\
\qed
\begin{lem}\lbl{fast-map}
The first return map is given by the 
\be\lbl{9.22}
\bar\xi=\mu\xi\left(1+\sigma r_1\right)+\sigma r_2 +o(\sigma)+O(\xi_0^2),
\ee
where Gaussian 
RVs $r_{1,2}$ are given by
\be\lbl{9.23}
r_1=-a(0)\int_0^\Tb \left[h_1(u)+B(\Tb,u)h_2(u)\right]dw_u,
 \quad r_2=\int_0^\Tb A(\Tb,u) h_2(u)dw_u.
\ee
\end{lem}
\pf:\;
From (\ref{9.5}), (\ref{9.7})-(\ref{9.9}), and (\ref{9.16}), we have 
\begin{eqnarray}\nonumber
\bar\xi &=& \xi_T= \xi_0 A(T) +\sigma \int_0^T A(T,s) h_2(s)dw_s +O(\sigma^2,\xi_0^2)\\
&=& \xi_0 A(T) +\sigma r_2 +O(\sigma^2,\xi_0^2), \lbl{9.24}
\end{eqnarray}
where $r_2$ is defined in (\ref{9.23}). The first term on the right
hand side of (\ref{9.24}) can be rewritten as follows
\begin{eqnarray}\nonumber
 A(T)&=&A(\Tb)A(\Tb+\sigma T^{(1)},\Tb) +o(\sigma)+O(\xi_0)=\mu
 \exp\left(\sigma a(0) T^{(1)}\right) +o(\sigma)\\
&=& \mu\left( 1-\sigma a(0) \theta_\Tb^{(1)}\right)
+o(\sigma)+O(\xi_0).\lbl{9.25}
\end{eqnarray}
Finally, we extract the expression for $\theta_\Tb^{(1)}$ from
(\ref{9.8}) and (\ref{9.9}):
\be\lbl{9.26}
\theta_\Tb^{(1)}=\int_0^\Tb \left[h_1(u)+B(\Tb,u)h_2(u)\right]dw_u.
\ee
Equations (\ref{9.24})-(\ref{9.26}) yield (\ref{9.22}) and (\ref{9.23}).
\qed
\begin{rem}\lbl{rem:gauss}
We close this section by observing that as follows from (\ref{9.23}) 
RV $r_1$ and $r_2$ are stochastic integrals of different deterministic functions, 
say $f(t)$ and $g(t)$ with respect to the same Brownian motion over 
the interval $[0,\mathcal T]$. Consequently, their joint distribution is bivariate normal with 0 mean vector and a covariance matrix that whose  diagonal entries are
$$\int_0^\mathcal T f^2(t)dt   \quad\mbox{and}\quad \int_0^\mathcal T g^2(t)dt,
$$ and the off diagonal entry is
$$\int_0^\mathcal T f(t)g(t)dt.$$
This is perhaps easiest to see by using Riemann representation of a stochastic integral (see e.g. \cite[Proposition~7.6]{steele}), basic properties of Brownian motion, and a fact that a random vector is multivariate normal if and only if any linear combination of its components is a normal RV.       
\end{rem}

\subsection{ The first return map for the slow variable}
Our next goal is to estimate the change of the slow variable, $y_t$, after
one cycle of oscillations of the fast subsystem for the following initial
conditions:
\be\lbl{6.1}
0<y_{bp}-y_0=O(1),\;x_0=\phi(0)+\xi_0\nu(0)\in \Sigma,\;\;\mbox{and}\;
\left|\xi_0\right|<\delta.
\ee
We denote the first return map for $y$ by
$$
\bar y= P(y,\xi_0), \quad\mbox{where}\quad P(y_0,\xi_0)=y_T,
$$
and $T$ is the first return time of the fast subsystem (see Definition \ref{return_time}).

\begin{lem}\lbl{slow}
The first return map for $y$
has the following form:
\be\lbl{6.2}
P(y,\xi)=y+\epsilon G(y) +\epsilon\sigma r_3 +\epsilon a\xi + o(\epsilon\sigma),
\ee
where
\be\lbl{6.3}
G(y)=\int_0^\Tb g\left(\phi(s), y\right)ds
\ee
and $r_3=N\left(0,O(1)\right)$ and $a$ is a constant independent of $\sigma$ and $\epsilon.$
\end{lem}
\begin{rem}
Recall that $\Tb$ and $\phi(\cdot)$ are functions of slow variable $y$ 
(see (\ref{1.4})).
To avoid using cumbersome notation we continue to suppress the dependence on $y$.
\end{rem}
\pf:\;
By (\ref{0.2}),
\be\lbl{6.4}
y_T=y_0+\epsilon\int_0^{T} g(x_s,y_0)ds +O(\epsilon^2),
\ee
where $x_s$ satisfies IVP (\ref{2.5}), (\ref{2.6}), and (\ref{9.1}).
Let $x=\phi(\theta)+\xi\nu(\theta)$ and denote
\be\lbl{N}
\tilde g(\theta,\xi,y):=g(x,y),\; g_0(s)=\tilde g(s,0),\; g_1(s)={\p\tilde g\over\p\theta} (s,0),\;
\mbox{and}\; g_2(s)={\p\tilde g\over\p\xi} (s,0).
\ee
Using (\ref{N}), we rewrite (\ref{6.4}) as
\be\lbl{6.5}
y_T= y_0+\epsilon\int_0^T\tilde g(\theta^{(0)}_s+\sigma\theta^{(1)}_s,\xi^{(0)}_s+\sigma\xi^{(1)}) +O(\epsilon\sigma^2).
\ee
Using the Taylor expansion for $\tilde g$ in (\ref{6.5}) and (\ref{9.4}), (\ref{9.5}) and 
(\ref{9.16}), from (\ref{6.5}) we derive
\begin{eqnarray}
\nonumber
y_T&=&y_0+ \eps \int_0^\Tb \left\{g_0(s)+g_1(s)\left[\xi_0B(s)+\sigma \theta^{(1)}_s\right]+
      g_2(s)\left[\xi_0 A(s)+\sigma\xi_s^{(1)}\right]\right\}ds\\
    &&+\int_{\Tb}^{\Tb-\xi_0B(\Tb)-\sigma \theta^{(1)}_\Tb} g_0(s)ds+o(\epsilon\sigma)+O(\epsilon\xi_0^2). \lbl{6.6}
\end{eqnarray}
We approximate the last integral on the right hand side of (\ref{6.6}) by
\be\lbl{6.7}
\int_{\Tb}^{\Tb-\xi_0B(\Tb)-\sigma \theta^{(1)}_\Tb}g_0(s)ds=-g_0(0)\left[ \xi_0 B(\Tb)+\sigma\theta^{(1)}\right] +
o(\sigma,\xi_0).
\ee
The combination of (\ref{6.6}) and (\ref{6.7}) implies (\ref{6.2}) with
\begin{eqnarray}\lbl{6.8}
a&=& \int_0^\Tb \left[g_1(s) B(s)+g_2(s)A(s)\right]ds-g_0(0)B(\Tb),\\
r_3&=& \int_0^\Tb \left[g_1(s) \theta_s^{(1)}+g_2(s)\xi_s^{(1)}\right]ds. \lbl{6.9}
\end{eqnarray}
\qed

\subsection{The exit problem}\lbl{sec:exit}
In the present subsection, we first combine the return maps derived for the slow and fast
subsystems to obtain the Poincare map for the full three-dimensional system. Next, we approximate
the Poincare map and the BA of the limit cycle $L(y_c)$ and characterize the
distribution of the exit times for the approximate problem. This distribution is then related to the
distribution of the number of spikes within bursting episodes. To approximate the Poincare map
we linearize it around the stable fixed point of the deterministic map corresponding to the 
limit cycle $L(y_c)$. Aside from the systematic derivation of the Poincare map in the previous
subsections, we offer no rigorous justification for substituting the nonlinear
Poincare map with its linear part in the analysis of the exit problem. 
While in general, such approximation may not be accurate,
we believe that for the present problem, the analysis of the linearized system captures the
statistics of the first exit times well for the following reason. In models of square wave
bursting the limit cycle generating spiking is often located close to the boundary 
of its BA (see Fig. \ref{f.4}b for a representative example). Therefore, before the trajectories 
leave the BA, they remain in a small neighborhood of the limit cycle, where the linear 
part of the vector field governs the dynamics. After these preliminary remarks, 
we turn to the derivation of the approximate problem and its analysis.

Lemmas \ref{fast-map} and \ref{slow} yield the asymptotic formulae for 
the first return map of the 
randomly perturbed system (\ref{0.1}) and (\ref{0.2}) 
in the normal coordinates (\ref{2.4}):
\begin{eqnarray}\lbl{6.01}
\xi_{n+1} &=&\mu\xi_n\left(1+\sigma r_{1,n}\right)+\sigma r_{2,n} +o(\sigma),\\
\lbl{6.02}
y_{n+1} &=&y_n+\epsilon G(y_n) +\epsilon\sigma r_{3,n} +\epsilon a\xi_n + o(\epsilon\sigma),\;
n=0,1,2,\dots,
\end{eqnarray}
where $(\xi_0, y_0)$ are given in (\ref{6.1}) and the expressions for $a$ and $r_{i,n},\; i=1,2,3$ are
are given in (\ref{9.23}),(\ref{6.8}), and (\ref{6.9}).
Recall that by (SS) (see Section 2), $G(y)$ has a simple zero at $y=y_c$ 
and $\lambda:=-G^\prime (y_c)>0$. Thus, $(0,y_c)$ is an attracting fixed point of the unperturbed
map (\ref{6.01}) and (\ref{6.02}) with $\sigma=0$. The linearization of (\ref{6.01}) and 
(\ref{6.02}) about $(0,y_c)$ yields
\begin{eqnarray}
\lbl{6.11a}
\xi_{n+1} &=&\mu\xi_n\left(1+\sigma \tilde r_{1,n}\right)+\sigma \tilde r_{2,n},\\
\lbl{6.12a}
\eta_{n+1} &=& \lambda \eta_n +\epsilon\sigma \tilde r_{3,n} +\epsilon a_2\xi_n, \quad n=0,1,2,\dots,
\end{eqnarray}
where $\eta=y-y_c$, $0<\lambda=1-\epsilon a_1$, and $0<\mu<1$. The distributions of the
RVs $r_{i,n},\; i=1,2,3$ depend on $y_n$, as both the upper bound of integration 
$\mathcal{T}$ and the integrands in (\ref{9.23}) and (\ref{6.9}) are smooth 
functions of $y$. The stochastic terms 
$\tilde r_{i,n},\; i=1,2,3$ in the linearized system are obtained by evaluating the expressions for  
$\tilde r_{i,n},\; i=1,2,3$ in (\ref{9.23}) and (\ref{6.9}) at $y=y_c$. 
Thus, $(\tilde r_{1,n}, \tilde r_{2,n},\tilde r_{3,n})$
are IID copies of a $N\left(0,\Sigma_3\right)$, where the entries of $\Sigma_3$ are $O(1)$ in a 
sense that they do not depend on any other parameters. Further, we approximate the BA 
of $L(y_c)$ by a cylindrical shell, so that in $(\xi,\eta)$ coordinate plane, 
it projects to 
$\Pi:=\left[ -\tilde h_\xi, h_\xi\right]\times \left[-\tilde h_\eta, h_\eta\right]$
for some $\tilde h_{\xi,\eta}>h_{\xi,\eta}>0$ independent of $\sigma>0$.
Each iteration of the Poincare map corresponds to a spike within a burst.
The burst terminates when the trajectory leaves the BA of $L(y_c)$.
Assuming that the linearization (\ref{6.11a}) and (\ref{6.12a}) and 
$\Pi$ provide suitable approximations 
for the Poincare map and the BA of $L(y_c)$ respectively, 
the distribution
of the number of spikes in one  burst can be approximated by the distribution of the first
exit times for the trajectories of (\ref{6.11a}) and (\ref{6.12a}) from $\Pi$:
\be\lbl{6.13}
\tau=\min\{\tau_\xi, \tau_\eta\},
\ee
where
$$
\tau_\xi =\inf_{n>0}\{ \xi_n>h_\xi\} \quad\mbox{and}\quad
\tau_\eta =\inf_{n>0}\{ \eta_n>h_\eta\}.
$$
We are now in a position to apply the the results of Section 3 to describe
the distribution of (\ref{6.13}).
By Theorem \ref{thm:2d-geom}, the distribution of $\tau$ is asymptotically geometric with parameter
\be\lbl{6.14}
p \approx{\sigma\over C\sqrt{2\pi}} e^{-\frac{C}{\sigma^2}}
\ee
for some $C>0$ independent of $\eps$ and $\sigma$. In the proof of Theorem~\ref{thm:2d-geom}, we studied
a class of $2D$ randomly perturbed maps that includes (\ref{6.11a}) and (\ref{6.12a}).
However, the distribution of $\tau$ is effectively determined by the first equation
(\ref{6.11a}),  i.e. by the
$1D$ first return map of the fast subsystem. 
This can be seen by observing that according to the approximations given at the end of proof of 
Theorem~\ref{thm:2d-geom} (see the arguments following (\ref{2nd}))
if $\epsilon>0$ is sufficiently small then 
$\tau_\xi\ll \tau_\eta$ and $\tau\sim\tau_\xi$. Thus, in type I models
the distribution of spikes in one burst is effectively determined by the
$1D$ first return map for the fast subsystem (\ref{6.11a}). In particular,
the statistics of the number of spikes in one burst does not depend on the
relaxation parameter $\epsilon>0$, provided the latter is sufficiently small.

\subsection{Type II model}
The derivation of the Poincare map for the type II models differs from 
the analysis in Sections 4.1-4.4 for type I
models only in some minor details. In this subsection, we comment on the necessary modifications
and state the final result. Recall that in contrast to type I models, in (\ref{0.1a}) and (\ref{0.2a}),
stochastic forcing enters the slow equation. As before, the initial condition is given by (\ref{6.1}).
On finite time intervals, solutions of the
IVP for (\ref{0.1a}) and (\ref{0.2a}) admit the following asymptotic expansions
\begin{eqnarray}\lbl{6.23}
x_t &=& x^{(0)}_t+\epsilon\sigma x^{(1)}_t+O\left( (\epsilon\sigma)^2\right),\\
\lbl{6.24}
y_t &=& y^{(0)}_t+\epsilon\sigma y^{(1)}_t+O\left( (\epsilon\sigma)^2\right).
\end{eqnarray}
where the first order corrections $x^{(1)}_t$ and $y^{(1)}_t$ are
Gaussian processes (cf. Theorem 2.2 \cite{FW}). Using (\ref{6.23}) and (\ref{6.24}),
we obtain the leading order approximation of the fast subsystem:
\be\lbl{6.25}
\dot x_t=f(x_t,y_0)+\epsilon\sigma {\p f(x_t^{(0)},y_0)\over \p y} y_t^{(1)} + o(\epsilon\sigma).
\ee
From this point, the derivation of the Poincare map follows the same lines as we described in detail
for type I models in Sections 4.1-4.4. We omit any further details and state the final result,
the linear approximation of the Poincare map for the present case:
\begin{eqnarray}
\lbl{6.26}
\xi_{n+1} &=&\mu\xi_n\left(1+\epsilon\sigma \tilde r_{1,n}\right)+\epsilon\sigma \tilde r_{2,n},\\
\lbl{6.27}
\eta_{n+1} &=&\lambda \eta_n +\epsilon\sigma \tilde r_{3,n} +\epsilon a_2\xi_n, \quad n=0,1,2,\dots,
\end{eqnarray}
As in the previous case, we are interested in the distribution of the first exit time $\tau$
(see (\ref{6.13})). To estimate the latter, we use the same argument as in the previous
subsection. 
This time the system is described by
\be\lbl{recII} \Theta_{n+1}=A_{n+1}\Theta_n+\sigma\epsilon G_{n+1},\quad n\ge1,
\ee
where
$A_n$ is as before  and $G_n=\left[\begin{array}{c}r_{2,n}\\ r_{3,n}\end{array}\right]
$. 
The presence of the factor $\epsilon$ in both components of $G_n$ leads to the following expression for the numerator of $p$ (see (\ref{limit})):
$$
\P\left(\frac{\mu X_1 r_1+r_2}{\sigma_{X_1}}>\frac{h_1-\mu X_1}{\epsilon\sigma\sigma_{X_1}},r_3>\frac{h_2- a_2 X_1}{\sigma}+\frac{\la(h_2-X_2)}{\epsilon\sigma},(X_1,X_2)\in B_h\right).
$$
This expression decays very fast as a function of $h_2-X_2$ and since $X_2$ has heavy tails it is approximated (up to inessential polynomial factors) by 
$$
\P\left(\frac{\mu X_1 r_1+r_2}{\sigma_{X_1}}>\frac{h_1-\mu X_1}{\epsilon\sigma\sigma_{X_1}},r_3>\frac{h_2- a_2 X_1}{\sigma},(X_1,X_2)\in B_h\right).
$$ 
We are now in the analogous situation to that encountered in (\ref{3rd}), except that the small parameter $\eps>0$ appears in the denominator of the other variable. As a consequence, this time we obtain that
$\tau_\xi\ll \tau_\eta$ for small
$\eps>0$.
Therefore, in contrast to type I models, the escape of a trajectory of (\ref{0.1a}) and (\ref{0.2a})
from $\mathcal{A}$ is dominated by the slow subsystem, i.e.,  $\tau=\tau_\eta$.

\section{Numerical example}
\begin{figure}
\begin{center}
% {\bf a}\epsfig{figure=type1_nsp.eps, height=2.0in, width=2.0in, angle=0}\qquad
{\bf a}\epsfig{figure=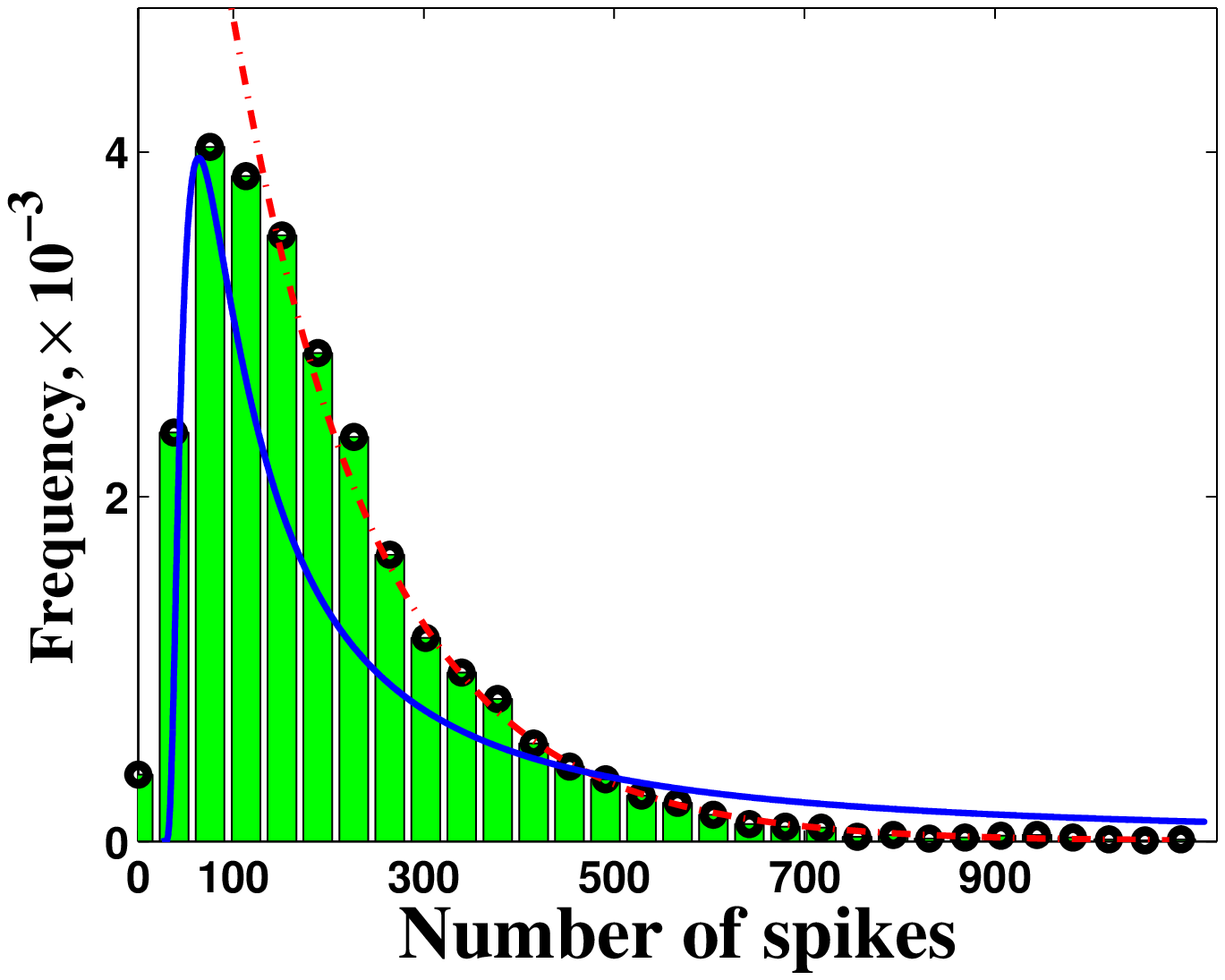, height=2.0in, width=2.0in, angle=0}\qquad
{\bf b}\epsfig{figure=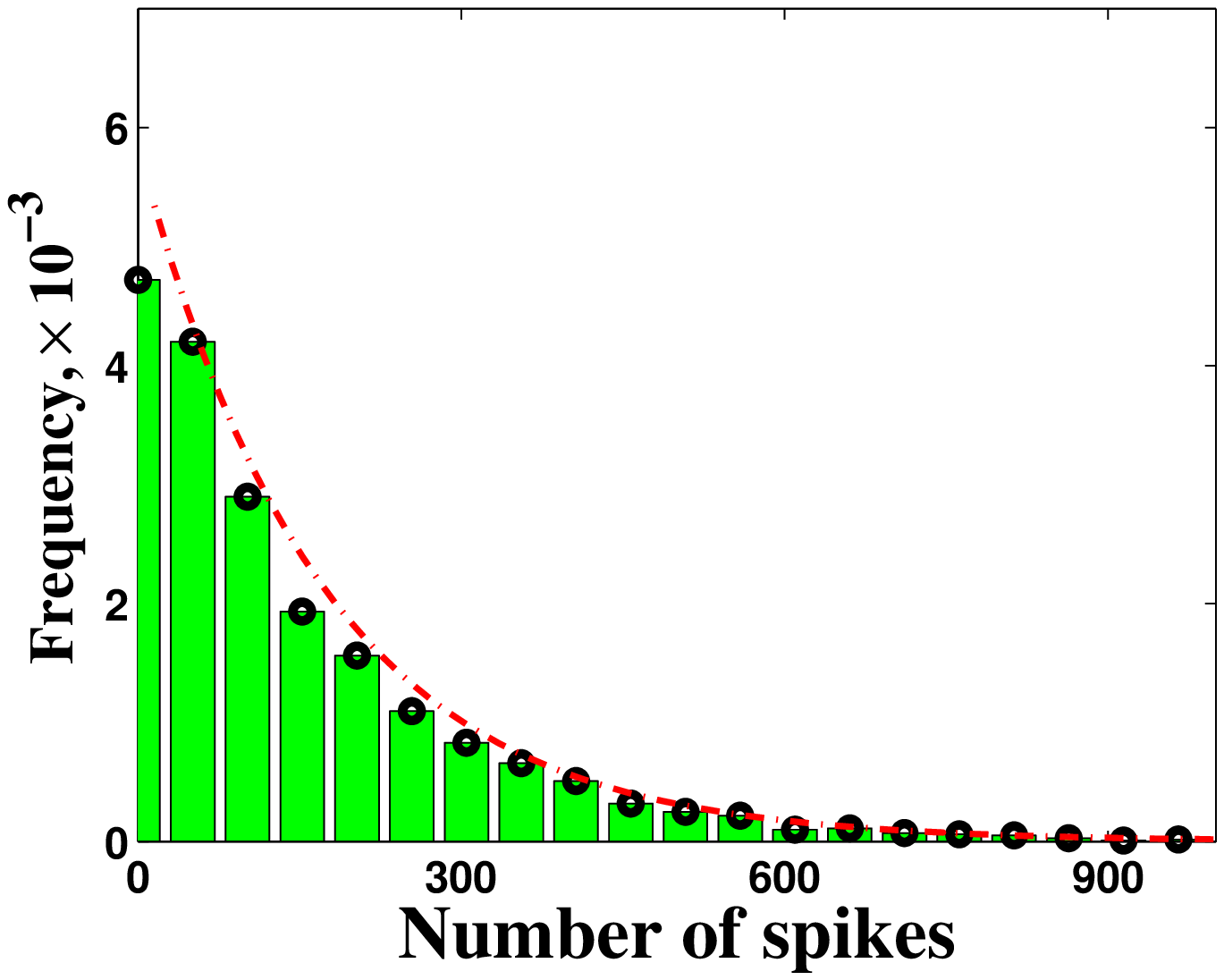, height=2.0in, width=2.0in, angle=0}
\end{center}
\caption{The histograms for the number of spikes in one burst.
The histograms computed for the type I model in  (a) and type II 
in (b) are normalized to approximate the corresponding PDFs. The tails of both functions are 
well approximated by the exponential densities with parameters $0.0067$ and $0.0125$ respectively. 
In (b) the exponential
distribution already gives a very good approximation for the number of spikes exceeding $10$. The region
of exponential behavior in (a) starts around  $n\sim 100$. 
In (a), we also plotted in solid blue line the shifted diffusive density  $\Psi_a(x-25), a\approx 10.8$. 
Although it is hard to claim a quantitative fit of the diffusive density and the data,
the qualitative similarity between the diffusive pdf $\Psi_a(x)$ and
the peak in the data in the range $n\sim 50-100$ is apparent.
The values of parameters
are $C=1$ $\left(\mu F cm^{-2}\right);$ $g_{NA}=20$, $g_K=10$, $g_{KM}=5$, 
$g_L=8$ $\left(mS cm^{-2}\right)$; $E_{Na}=60$, $E_K=-90$, 
$E_L=-80$ $\left(mV\right)$; 
$a_m=-20$, $a_n=-25$, $a_y=-10$ $\left(mV\right)$;
$b_m=15$, $b_n=5$, $b_y=5$; $\tau_n=0.152$, $\tau_y=20$ $\left(ms^{-1}\right)$,
$I=5 pA$, and $\sigma=1$.
}
\label{f.11}
\end{figure}

\setcounter{equation}{0}
In the present section, we illustrate the statistical regimes identified in this study with numerical
simulations of a conductance based model of a neuron in the presence of noise. 
To this end,  we use a three variable model of a
bursting neuron introduced by Izhikevich \cite{IZH07}. The model dynamics is driven by the interplay
of the three ionic currents: persistent sodium, $I_{NaP}$,
the delayed rectifier, $I_K$, a slow  potassium $M$-current, $I_{KM}$, and a passive leak current
$I_L$. The following system of three differential equations describes the dynamics
of the membrane potential, $v$, and two gating variables $n$ and $y$:
\begin{eqnarray}\lbl{10.1}
C\dot v &=&F(v,n,y),\\\lbl{10.2}
\tau_n \dot n &=& n_\infty(v)-n,\\
\lbl{10.3}
\tau_y \dot y &=& y_\infty(v)-y,
\end{eqnarray}
where $F(v,n,y)=-g_{NaP}m_\infty(v)(v-E_{NaP})-g_Kn(v-E_K)-g_{KM} y(v-E_{K})-g_L(v-E_L)+I$;
$g_s$ and $E_s$, are the maximal conductance
and the reversal potential of $I_s$, $s \in \left\{NaP, K, KM, L\right\}$, respectively; and
$I$ is the applied current. The time constants $\tau_n$ and $\tau_y$ determine the rates of
activation in the populations of $K$ and $KM$ channels. The
steady-state functions are defined by
$
s_\infty (v) =\left(1+\exp\left(\frac{a_s-v}{b_s}\right)\right)^{-1},\; s\in\left\{m,n,y\right\}.
$
The parameter values are given in the caption to Fig. \ref{f.11}. 
This completes the description of the deterministic model. 
The random perturbation is used in the form of white noise, $\sigma\dot w_t$ and is added to 
the first equation (\ref{10.1}) for type I model or to the third one (\ref{10.3}) for type II
model. After suitable rescaling, these models can be put in the nondimensional form
(\ref{0.1}), (\ref{0.2}) or (\ref{0.1a}), (\ref{0.2a}). The separation of the timescales in the
nondimesional models (i.e. small $\eps>)$) is the result of the presence of the disparate 
time constants $\tau_h\gg \tau_n$ in the original model (see caption to Fig. \ref{f.11}).

The parameters of the deterministic system are chosen so that it has a limit cycle located as shown in Fig. 
\ref{f.1}c. In the presence of small noise the system generates bursting. 
In each  numerical experiment, we integrated
the randomly perturbed system using the Euler-Maruyama method \cite{higham01} until
it generated $5,000$ bursts. We used these data to estimate the probability density for the number
of spikes within one burst.
In Fig. \ref{f.11}, we plot
the histograms for the number of spikes in one burst for type I and type II models. The histograms in
Fig. \ref{f.11} are scaled to approximate the probability density function (PDF) for the number of spikes 
in one burst. Both PDFs shown Fig. \ref{f.11}a,b have distinct exponential tails as expected for 
the asymptotically geometric RVs. 
Note that the distribution in Fig. \ref{f.11}a
fits well with the geometric distribution for $N>100$, while in Fig. \ref{f.11}b the geometric distribution
fits the data almost on the entire domain $N>10$. 
In addition, the peak in the histogram in Fig.~\ref{f.11}a is reminiscent of the PDF characteristic for 
the diffusive escape (see Fig.~\ref{f.3}). For comparison, we plotted a slightly shifted diffusive
PDF, $\Psi_a(x),\;a=10.8$ in Fig.~\ref{f.11}a. Matching the data and $\Psi_a$ is a delicate matter,
because it is not clear how wide is the range of $n$, to which the estimates of Theorem~\ref{eps=0}
apply. Nonetheless, the qualitative similarity of the peak in the histogram in the range $n\sim 50-100$
and the diffusive PDF is apparent. We repeated these
numerical experiments for a few other sets of parameters and found qualitatively similar results. 

Collecting the  statistical data shown in Fig. \ref{f.11} requires integrating the system over very 
long intervals of time, for which it would be hard to justify the accuracy of the Euler-Maruyama method.
However, capturing the statistical features of the dynamical patterns does not require having an accurate
solution on the entire interval of time, because they are determined by the discrete dynamics of the first 
return map. The iterations 
of the latter are expected to be insensitive to the  numerical noise as suggested by the analysis of the 
randomly perturbed maps in Section 3. Therefore, we only need to have accurate numerical solutions 
on the time intervals comparable with the typical periods of the fast oscillations. 
This is easy to achieve with the  Euler-Maruyama method.
We repeated these numerical experiments using the second order Runge-Kutta method 
and obtained very similar results.
These informal arguments form the rationale for using the above   numerical 
scheme. The rigorous justification of the numerics is beyond the scope of this paper.

\noindent
{\bf Acknowledgments.}
This work was partially supported by NSF grant IOB 0417624 (to GM) and 
NSA grant MSPF-04G-054 (to PH).

\section*{Apendix}
\def\theequation{A.\arabic{equation}}
\renewcommand{\theequation}{A.\arabic{equation}}
\setcounter{equation}{0}

In this appendix, we provide the details of the derivation of (\ref{9.6})-(\ref{9.9}),
which were omitted in the main part of the paper. 

To derive (\ref{9.6}) and (\ref{9.7}), we first note that $\theta_t^{(0)}$ is a monotonic function
on $[0,\bar t]$, provided $\delta>0$ is sufficiently small. Thus,
$$
{d\xi^{(0)}\over d\theta^{(0)}}=a(\theta^{(0)})\xi^{(0)}+O(\xi_0^2),
$$
and
\be\lbl{a.1}
\xi^{(0)}(\theta^{(0)})=\xi_0 A(\theta^{(0)})+O(\xi_0^2).
\ee
By plugging in (\ref{a.1}) into (\ref{9.10}), we have 
\be\lbl{a.2}
\dot\theta_t^{(0)}=1+b_1(\theta_t^{(0)})\xi_0 A(\theta^{(0)}).
\ee
By Gronwall's inequality, 
\be\lbl{a.3}
\theta_t^{(0)}=\psi_t+O(\xi_0^2), \; t\in [0, \bar t],
\ee
where $\psi_t$ solves
\be\lbl{a.4}
\dot\psi_t^{(0)}=1+\xi_0 b_1(t)A(t), \; \psi_0=0.
\ee
The combination of (\ref{a.1}), (\ref{a.3}), and (\ref{a.4}) implies (\ref{9.7}).

We next turn to IVP (\ref{9.13}), (\ref{9.14}) and (\ref{9.7}). Let $U(t,\xi_0)$
denote the principal matrix solution of the homogeneous system 
\be\lbl{a.5}
\dot z_t=\Lambda(t,\xi_0) z_t.
\ee
Then the solution of (\ref{9.13}) and (\ref{9.14}) is given by
\be\lbl{a.6}
z_t=\int_0^t U(t,s,\xi_0)h\left(\theta^{(0)}_s,\xi_s^{(0)}\right)dw_s=
\int_0^t U(t,s)h\left(s,0\right)dw_s + O(\xi_0),\; t\in [0,\bar t],
\ee
where
\be\lbl{a.7}
U(t,s,\xi_0)=U(t,\xi_0)U^{-1}(s,\xi_0)\quad\mbox{and}\quad U(t,s)=U(t,s,0).
\ee
Finally, by integrating (\ref{a.5}) with $\xi_0=0$ and  appropriate initial 
conditions, one computes
\be\lbl{a.8}
U(t,0)=\left(\begin{array}{cc}
                                1 & B(t)\\
                                0 & A(t) 
             \end{array}
       \right).
\ee
The expression for $U(t,s)$ in (\ref{9.9}) follows from (\ref{a.7}) and (\ref{a.8}).

\vfill
\break

\end{document}